\documentclass[aps,prl,longbibliography,reprint,showpacs,floatfix,superscriptaddress]{revtex4-1}
\usepackage{graphicx}
\usepackage{amsmath,amssymb,amsfonts}
\usepackage[english]{babel}
\usepackage[bookmarks=false]{hyperref}

\begin{document}
\title{Magnetic properties of single crystalline itinerant ferromagnet AlFe$_{2}$B$_{2}$}
\author{Tej N. Lamichhane} 
\affiliation{Ames Laboratory, U.S. DOE, Ames, Iowa 50011, USA}
\affiliation{Department of Physics and Astronomy, Iowa State University, Ames, Iowa 50011, USA}
\author{Li Xiang}
\affiliation{Ames Laboratory, U.S. DOE, Ames, Iowa 50011, USA}
\affiliation{Department of Physics and Astronomy, Iowa State University, Ames, Iowa 50011, USA}
\author{Qisheng Lin}
\affiliation{Ames Laboratory, U.S. DOE, Ames, Iowa 50011, USA}
\author{Tribhuwan Pandey}
\affiliation{Materials Science and Technology Division, Oak Ridge National Laboratory, Oak Ridge, Tennessee 37831, USA}
\author{David S. Parker}
\affiliation{Materials Science and Technology Division, Oak Ridge National Laboratory, Oak Ridge, Tennessee 37831, USA}
\author{Tae-Hoon Kim}
\affiliation{Ames Laboratory, U.S. DOE, Ames, Iowa 50011, USA}
\author{Lin Zhou}
\affiliation{Ames Laboratory, U.S. DOE, Ames, Iowa 50011, USA}
\author{Matthew J. Kramer}
\affiliation{Ames Laboratory, U.S. DOE, Ames, Iowa 50011, USA}
\author{Sergey L. Bud'ko}
\affiliation{Ames Laboratory, U.S. DOE, Ames, Iowa 50011, USA}
\affiliation{Department of Physics and Astronomy, Iowa State University, Ames, Iowa 50011, USA}  
\author{Paul C. Canfield} 
\affiliation{Ames Laboratory, U.S. DOE, Ames, Iowa 50011, USA}
\affiliation{Department of Physics and Astronomy, Iowa State University, Ames, Iowa 50011, USA}
\begin{abstract}
   Single crystals of AlFe$_{2}$B$_{2}$ have been grown using the self flux growth method and then measured the structural properties, temperature and field dependent magnetization,  and temperature dependent electrical resistivity at ambient  as well as high pressure. The Curie temperature of AlFe$_{2}$B$_{2}$ is determined to be $274$~K.  The measured saturation magnetization  and  the effective moment for paramagnetic Fe-ion  indicate the itinerant nature of the magnetism with a Rhode-Wohlfarth ratio $ \frac{M_{C}}{M_{sat}}\approx 1.14$. Temperature dependent resistivity measurements under hydrostatic pressure shows that transition temperature \textit{T$_C$} is suppressed down to 255 K for $p = 2.24$~GPa pressure with a suppression rate of $\sim -8.9$~K/GPa. The anisotropy fields and magnetocrystalline anisotropy constants are in reasonable agreement with density functional theory calculations.     
\end{abstract}

\maketitle

\section{Introduction}
In recent years, AlFe$_{2}$B$_{2}$ has attracted a growing research interest as a rare-earth free ferromagnet that might have potential as a  magneto-caloric material~\cite{Tanetal.AlFe2B2, Cedervall2016}. It is a layered material that has been identified as an itinerant ferromagnet \cite{ElMassalamiAlFe2B2}. AlFe$_2$B$_2$ was first reported by Jeitschko\cite{WJeitschkoFe2AlB2} and independently  by Kuz'ma  and Chaban \cite{Kuzma1969} in 1969. AlFe$_{2}$B$_{2}$ crystallizes in an orthorhombic structure with space group \textit{Cmmm} (Mn$_2$AlB$_2$ structure type). The Al atoms located in 2\textit{a} crystallographic position (0,0,0) form a plane which alternately stacks with Fe-B slabs formed by Fe atoms; located at 4\textit{j}(0, 0.3554, 0.5) and B atoms located at 4\textit{i} (0,0.1987,0) positions \cite{QianhengDuMndopedAlFe2B2}. A picture of unit cell for  AlFe$_{2}$B$_{2}$ is shown in FIG.~\ref{TEM}(a). AlMn$_{2}$B$_{2}$ and AlCr$_{2}$B$_{2}$ are the other two known iso-structural transition metal  compounds. Among these 3 members only AlFe$_{2}$B$_{2}$ is ferromagnetic; however the reported magnetic parameters for AlFe$_{2}$B$_{2}$ show a lot of variation ~\cite{ElMassalamiAlFe2B2, Tanetal.AlFe2B2, QianhengDuMndopedAlFe2B2, Cedervall2016, HirtAlFe2B2Co}. A good summary of all these variations is presented tabular form in a very recent literature~\cite{BARUA2018}.

 For example, the Curie temperature of this material is reported to fall within a window of 274 - 320 K depending up on the synthesis route. Initial work indicates that, the Curie temperature of AlFe$_{2}$B$_{2}$ was $320$~K \cite{ElMassalamiAlFe2B2}. The Curie temperature of Ga-flux grown AlFe$_{2}$B$_{2}$ was reported to be $307$~K and for arcmelted, polycrystalline samples it was reported to be $282$~K \cite{Tanetal.AlFe2B2}. The Curie temperature for annealed, melt-spun ribbons was reported to be $312$~K \cite{QianhengDuMndopedAlFe2B2}. A M{\"o}ssbauer study on arc-melted and annealed sample has reported the Curie temperature of $300$~K \cite{Cedervall2016}. At the lower limit, the Curie temperature of spark plasma sintered AlFe$_{2}$B$_{2}$ was reported to be $274$~K \cite{HirtAlFe2B2Co}. The reported saturation magnetic moment also manifests up to 25\% variation from the theoretically predicted saturation moment of 1.25~$\mu_{B}$/Fe. The first reported saturation magnetization and effective moment values for AlFe$_{2}$B$_{2}$ were 1.9(2)~$\mu_{B}$/f.u. at $4.2$~K and 4.8~$\mu_{B}$/Fe respectively \cite{ElMassalamiAlFe2B2}. Recently, Tan et al. has reported the saturation magnetization of 1.15~$\mu_{B}$/Fe and 1.03~$\mu_{B}$/Fe for before and after the HCl etching of an arcmelted sample \cite{Tanetal.AlFe2B2}. The lower saturation moment, after the acid etching, suggested either the inclusion of Fe-rich magnetic impurities in the sample or degradation of the sample with acid etching. Recently, a study pointed out that the content of impurity phases decreases with excess of Al in the as cast alloy and by annealing~\cite{Levin2018}. The main reason for the variation in reported magnetic parameters is the difficulty in preparing pure single crystal, single phase AlFe$_{2}$B$_{2}$ samples. To this end, detailed measurements on single phase, single crystalline samples will provide unambiguous magnetic parameters and general insight into AlFe$_{2}$B$_{2}$.\\*
In this work, we investigated the magnetic and transport properties of self-flux grown single crystalline  AlFe$_{2}$B$_{2}$. We report single crystalline  structural, magnetic and transport properties of AlFe$_{2}$B$_{2}$. We find that AlFe$_{2}$B$_{2}$ is an itinerant ferromagnet with $ \frac{M_{C}}{M_{sat}}\approx 1.14$ and the Curie temperature is initially linearly suppressed with hydrostatic pressure at rate of $\frac{dT_C}{dp}\sim -8.9$~K/GPa.  The magnetic anisotropy fields of  AlFe$_{2}$B$_{2}$ are $\sim 1$~T along [010] and $\sim 5$~T along [001] direction. The first magneto-crystalline anisotropic constants ($K_1$s) at base temperature are determined to be $K_{010} \approx 0.23 MJ/m^3$ and $K_{001} \approx 1.8 MJ/m^3$ along [010] and [001] directions respectively (The subscript 1 is dropped for simplicity.).
 \section{Experimental Details}
\subsection{Crystal growth}
Single crystalline samples were prepared using a self-flux growth technique\cite{CanfieldFisk}. First we confirmed that our initial stochiometry Al$_{50}$Fe$_{30}$B$_{20}$ was a single phase liquid at $1200~^\circ$C. Starting composition Al$_{50}$Fe$_{30}$B$_{20}$ with elemental Al (Alfa Aesar, 99.999\%), Fe (Alfa Aesar, 99.99\%) and B (Alfa Aesar, 99.99\%) was arcmelted under an Ar atmosphere at least 4 times. The ingot was then crushed with a metal cutter and put in a fritted alumina crucible set~\cite{Canfieldfrittedcrucible} under the partial pressure of Ar inside an amorphous SiO$_{2}$ jacket for the flux growth purpose. The growth ampoule was heated to $1200~^\circ$C over 2-4 h and allowed to homogenize for 2 hours. The ampoule was then placed in a centrifuge and all liquid was forced to the catch side of crucible. Given that all of the melt was collected in catch crucible, this confirms that Al$_{50}$Fe$_{30}$B$_{20}$ is liquid at $1200~^\circ$C.

\indent
Knowing that the arcmelted  Al$_{50}$Fe$_{30}$B$_{20}$ composition exists as a homogeneous melt at $1200~^\circ$C , the cooling profile was optimised as following. 
The homogeneous melt at $1200~^\circ$C was cooled down to $1180~^\circ$C over 1 h and slowly cooled down to $1080~^\circ$C over 30 h at which point the crucible limited, plate-like crystals were separated from the remaining flux using a centrifuge. The large plate-like crystals had some Al$_{13}$Fe$_{4}$ impurity phase on their surfaces which was removed with dilute HCl etching \cite{HirtAlFe2B2Co}. The as-grown single crystals are shown in the  insets of FIG.~\ref{SCsurface-XRD}(a).

\begin{figure}[!ht]
\begin{center}
\includegraphics[width=7cm]{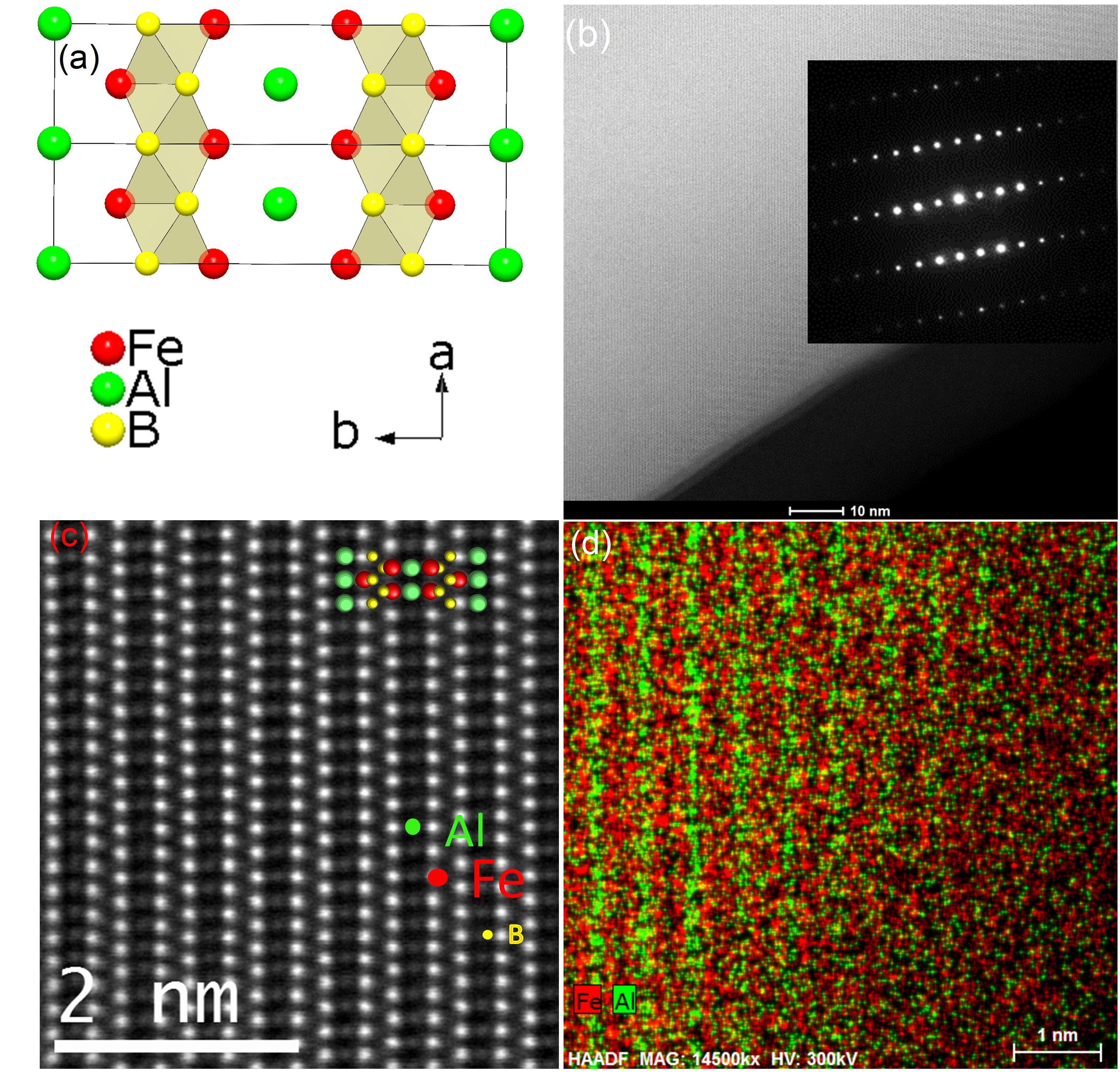}
\caption{(a) AlFe$_2$B$_2$ unit cell (b) HAADF STEM image shows uniform chemistry of the AlFe$_{2}$B$_{2}$  crystal. The inset is a corresponding selected-area electron diffraction pattern. (c) High resolution HAADF STEM image of AlFe$_{2}$B$_{2}$  taken along [101] zone axis along with projection of a unit cell represented with Fe (red), Al(green) and B (yellow) spheres. The structural pattern of Al and FeB slab layers are also visible in unit cell shown in pannel (a). (d) EDS elemental mapping without account of B scattering effect where green stripes are Al and red stripes are Fe distributions.}
\label{TEM}
\end{center}
\end{figure}

\subsection{Characterization and physical properties measurements}

The crystal structure of AlFe$_{2}$B$_{2}$ was characterized with both single crystal X-ray diffraction (XRD) and powder XRD. The single crystal XRD data were collected within a $4^\circ$-$29^\circ$ angle value of 2$\theta$ using Bruker Smart APEX II diffractometer with graphite-monochromatized Mo-K$_{\alpha}$ radiation source ($\lambda$ = 0.71073 $\AA$). The powder diffraction data were collected using a Rigaku MiniFlex II diffractometer with Cu-K$_{\alpha}$ radiation. The acid etched AlFe$_{2}$B$_{2}$ crystals were ground to fine powder and spread over a zero background, Si-wafer sample holder with help of a thin film of Dow Corning high vacuum grease. The diffraction intensity data were collected within a 2$\theta$ interval of $5^\circ$-$100^\circ$ with a fixed dwelling time of 3 sec and a step size of  $0.01^\circ$.

\indent
The as-grown single crystalline sample was examined with a transmission electron microscopy to obtain High-angle-annular-dark-field (HAADF) scanning transmission electron microscopy (STEM) images, corresponding selected-area electron diffraction pattern and high resolution HAADF STEM image of AlFe$_{2}$B$_{2}$ taken under [101] zone axis.

\indent
The anisotropic magnetic measurements were carried out in a Quantum Design Magnetic Property Measurement System (MPMS) for 2 K $\leq T \leq $ 300 K and a Versalab Vibrating Sample Magnetometer (VSM) for 50 K$\leq T \leq $700~K. 

\indent
The temperature dependent resistivity of AlFe$_2$B$_2$ was measured in a standard four-contact configuration, with contacts prepared using silver epoxy. The excitation current was along the crystallographic a-axis. AC resistivity measurement were performed in a Quantum Design Physical Property Measurement System (PPMS) using 1 mA; 17 Hz excitation, with a cooling at a rate of 0.25 K/min. A Be-Cu/Ni-Cr-Al hybrid piston-cylinder cell similar to the one described in Ref. \onlinecite{Budko1984} was used to apply pressure.  Pressure values at the transition temperature $T_C$ were estimated by linear interpolation between the room temperature pressure $p_{300\textrm{K}}$ and low temperature pressure $p_{T\leq90\textrm{K}}$ values\cite{Thompson1984,Torikachvili2015}. $p_{300\textrm{K}}$ values were inferred from the $300$ K resistivity ratio $\rho(p)/\rho(0\,\textrm{GPa})$ of lead\cite{Eiling1981} and $p_{T\leq90\textrm{K}}$ values were inferred from the $T_{c}(p)$ of lead\cite{Bireckoven1988}. Good hydrostatic conditions were achieved by using a 4:6 mixture of light mineral oil:n-pentane as a pressure medium; this mixture solidifies at room temperature in the range $3-4$ GPa, i.e., well above our maximum pressure\cite{Budko1984,Kim2011,Torikachvili2015}.

\section{Experimental results}

\subsection{Structural characterization}
The HAADF STEM image along with selected area diffraction pattern in the inset and high resolution HAADF STEM image of AlFe$_{2}$B$_{2}$ taken under [101] zone axis and EDS Al-Fe elmental mapping are presented in pannels (b), (c) and (d) of FIG.~\ref{TEM}. Taken together they strongly suggest the uniform chemical composition of AlFe$_{2}$B$_{2}$ through out the sample.

\indent

\begin{figure*}[!ht]
\begin{center}
\includegraphics[width=16cm]{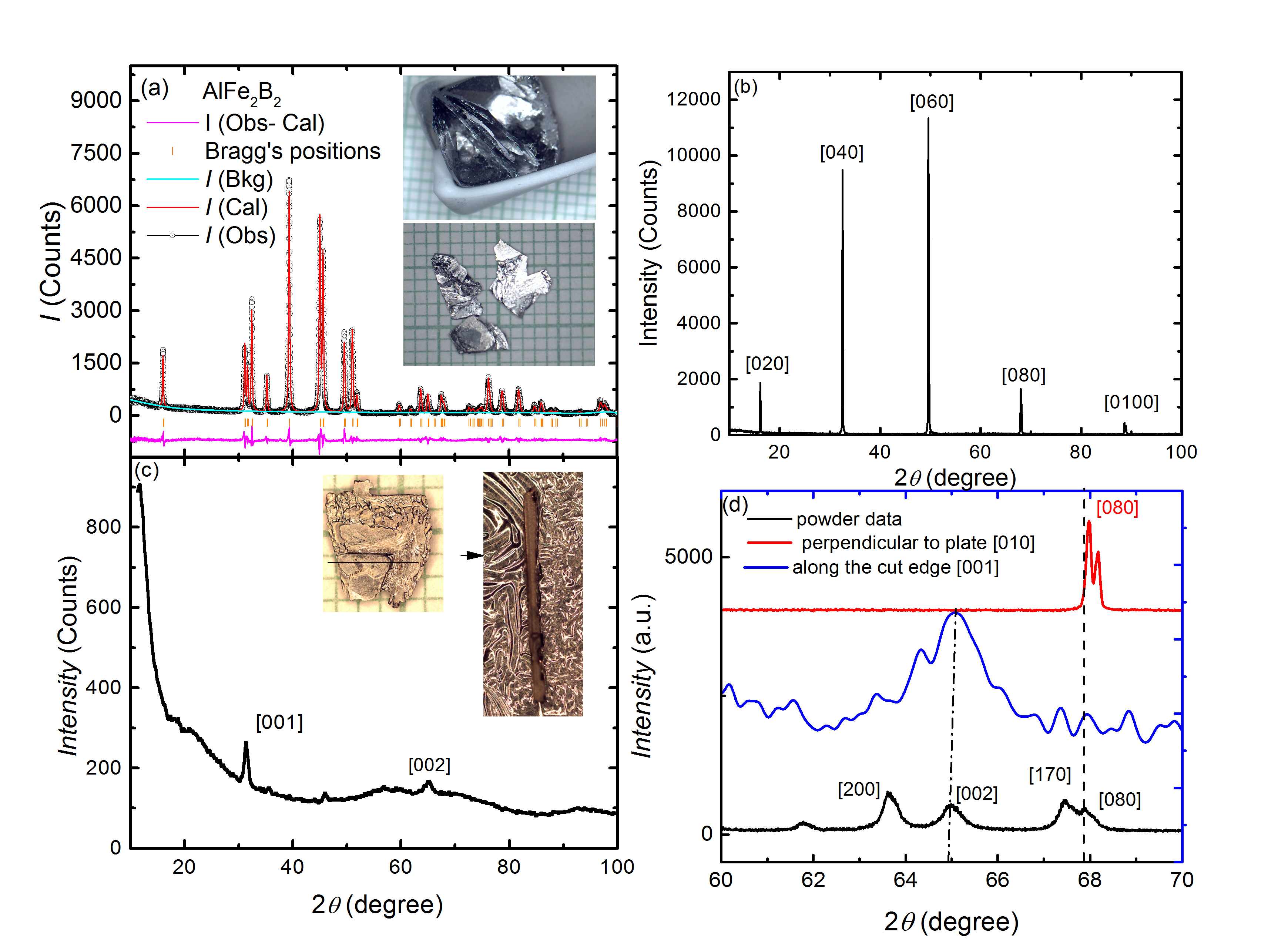}
\caption{(a) Powder XRD for AlFe$_{2}$B$_{2}$. \textit{I}(Obs), \textit{I}(Cal) and \textit{I}(Bkg) stands for experimental powder diffraction, Rietveld refined and instrumental background data. The green vertical lines represent the Bragg reflection peaks and the \textit{I}(Obs-Cal) is the differential intensity between \textit{I}(Obs) and \textit{I}(Cal). The upper inset picture shows the crucible limited growth nature of AlFe$_{2}$B$_{2}$. The lower inset picture is the pieces of as grown plate-like crystals. (b) Monochromatic XRD pattern from the plate surface of AlFe$_{2}$B$_{2}$. (c) Monochromatic XRD pattern from cut surface [001] collected using Bragg-Brentano geometry. The left inset photo shows the as grown AlFe$_{2}$B$_{2}$ crystal. The right inset picture is the photograph of the cut section of the crystal parallel to (001) plane. The middle unidentified peak might be due to a differently oriented shard of cut AlFe$_{2}$B$_{2}$ crystal. (d) Comparison of the monochromatic surface XRD patterns from (b) and (c) with powder XRD pattern from (a) within extended 2$\theta$ range of 60 - 70$^\circ$ to illustrate the identification scheme of the crystallographic orientation.}
\label{SCsurface-XRD}
\end{center}
\end{figure*}

The crystallographic solution and parameters refinement on the single crystalline XRD data was performed using SHELXTL program package \cite{SHELXTL}. The Rietveld refined single crystalline data are presented in TABLES~\ref{tbl:crystaldata}, and~\ref{tbl:atomcoord}. Using the atomic coordinates from the crystallographic information file obtained from single crystal XRD data, powder XRD data were Rietveld refined with $R_P$ = 0.1 using General Structure Analysis System~\cite{Toby:hw0089} (FIG.~\ref{SCsurface-XRD} (a)). The lattice parameters from the powder XRD are: a = 2.920(4)~$\AA$, b = 11.026(4)~$\AA$ and c = 2.866(7)~$\AA$ which are in reasonable agreement with the single crystal data analysis values.
 
\begin{table}[!htbp]
\begin{center}
\caption{\label{tbl:crystaldata}Crystal data and structure refinement for AlFe$_{2}$B$_{2}$.}
\begin{tabular}{|l|l|}
\hline
Empirical formula & AlFe$_{2}$B$_{2}$ \\
Formula weight & 160.3 \\
Temperature & $293(2)$~K \\
Wavelength & $0.71073$~\AA \\
Crystal system, space group & Orthorhombic,  \textit{Cmmm} \\
Unit cell dimensions & a=2.9168(6)~\AA \\
 & b = 11.033(2)~\AA \\
 & c = 2.8660(6)~\AA \\
Volume &  92.23(3) $10^3$ \AA$^3$ \\
Z, Calculated density & 2,  5.75 g/$cm^3$ \\
Absorption coefficient & 31.321 mm$^{-1}$ \\
F(000) & 300 \\
$\theta$ range ($^\circ$) & 3.693 to  29.003 \\
Limiting indices & $-3\leq h \leq3$\\
& $-14\leq k\leq 14$\\
& $-3\leq l \leq 3$ \\
Reflections collected & 402\\
Independent reflections & 7 [R(int) = 0.0329] \\
Absorption correction & multi-scan, empirical \\
Refinement method & Full-matrix least-squares \\
& on F$^2$ \\
Data / restraints / parameters & 74 / 0 / 12 \\
Goodness-of-fit on $F^2$ & 1.193 \\
Final R indices [I$>2\sigma$(I)] & $R1 = 0.0181$, $wR2 = 0.0467$ \\
R indices (all data) & $R1 = 0.0180$, $wR2 = 0.0467$\\
Largest difference peak and hole & 0.679  and -0.880 e.\AA$^{-3}$\\ \hline
\end{tabular}
\end{center}
\end{table}

\begin{table}[!htbp]
\begin{center}
\caption{\label{tbl:atomcoord}Atomic coordinates and equivalent isotropic displacement parameters (A$^2$) for AlFe$_{2}$B$_{2}$. U(eq) is defined as one third of the trace of the orthogonalized U$_{ij}$ tensor.}
\begin{tabular}{| p{0.6cm} | p{1.2cm} | p{1.2cm} | p{1.8cm} | p{1.2cm} |p{1.8cm}|}
\hline
atom & Wyckoff site & x & y & z & U$_{eq}$ \\ \hline
Fe & 4(\textit{j}) & 0.0000 & 0.3539(1) & 0.5000 & 0.006(6) \\ \hline
Al & 2(\textit{a}) & 0.0000 &  0.0000 & 0.0000 & 0.006(7) \\ \hline
B & 4(\textit{i}) &  0.0000 & 0.2066(5) & 0.0000 & 0.009(7) \\ \hline
\end{tabular}
\end{center}
\end{table}

To confirm the crystallogrpahic orientation of the AlFe$_{2}$B$_{2}$ crystals, monochromatic Cu-$K_{\alpha}$ XRD data were collected from the flat surface of the crystals and found to be \{020\} family as shown in FIG.~\ref{SCsurface-XRD}~(b), i.e. the [010] direction is perpendicular to the plate. However finding a thick enough, flat, as grown facet with [100] and [001] direction was made difficult by the thin, sheet-like morphology of the sample and its crucible limited growth nature. A [001] facet was cut out of large crucible limited crystal as shown in the inset of FIG.~\ref{SCsurface-XRD}(c). The monochromatic  Cu-$K_{\alpha}$  XRD  pattern scattered from the cut surface confirms the [001] direction displaying the [001] and [002] peaks (FIG.~\ref{SCsurface-XRD}~(c)). To better illustrate the crystallographic orientations, powder XRD, and monochromatic surface XRD patterns from the  plate surface and cut edge are plotted together in FIG.~\ref{SCsurface-XRD}~(d). This plot clearly identifies that direction perpendicular to the plate is [010]  and cut edge surface is (001). Slight displacement of the surface XRD peaks is the result of the sample height in the Bragg Brentano geometry. The splitting of [080] peak is observed by distinction of  Cu-$K_{\alpha}$  satellite XRD patterns usually observed at high diffraction angles.

\begin{figure*}[!h]
\begin{center}
\includegraphics[width=18cm]{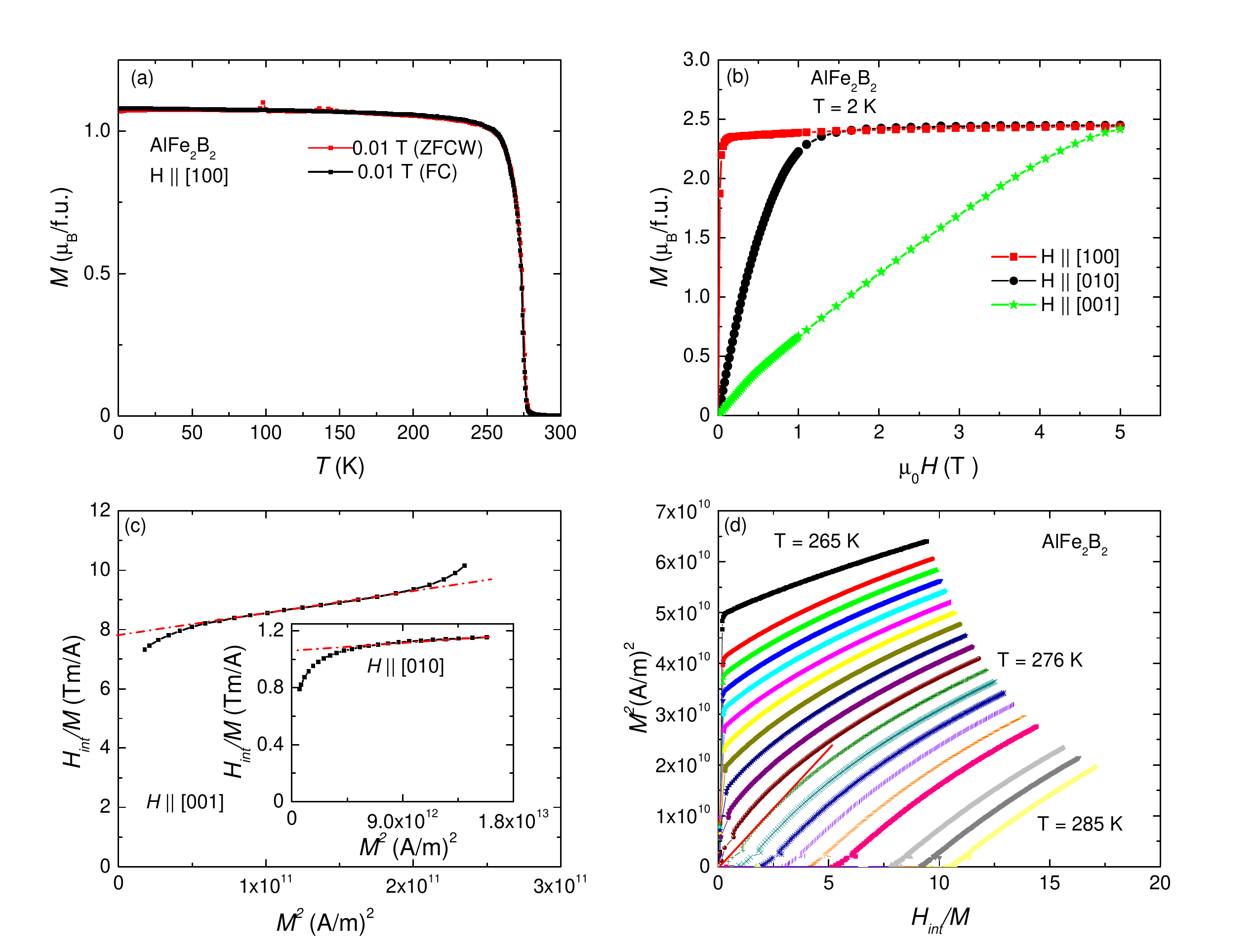}
\caption{(a) Temperature dependent magnetization with 0.01~T applied field along   [100] direction (b)) Field dependent magnetization along principle directions at $2$~K. [100] is the easy axis with smallest saturating field , [010] is the intermediate axis with 1 T anisotropy and [001] is the hardest axis with $\sim 5$~T anisotropy field. (c) Sucksmith-Thompson plot for \textit{M(H)} data along  [001] direction (and along [010] in the inset) to estimate the magneto-crystalline anisotropy constants. The red dash-dotted line is the linear fit to the hard axes isotherms at high field region ($>3$~T) whose Y-intercept is used to estimate the anisotropy constant $K$. (d) Arrott plot obtained with easy axis isotherms within the temperature range of 265-285 K at a step of 1~K. The straight line through the origin is the tangent to the isotherm corresponding to the transition temperature.}
\label{Combinedgraph}
\end{center}
\end{figure*}

\section{Magnetic properties}
The anisotropic magnetization data were measured using a sample with known crystallographic orientation. The temperature dependent magnetization \textit{M(T)} data along [100] axis is presented in FIG.~\ref{Combinedgraph}(a). Both the Zero Field Cooled Warming (ZFCW) and Field Cooled(FC) \textit{M(T)} data  are almost overlapping for 0.01 T applied field. The \textit{M(T)} data suggest a Curie temperature (\textit{T}$_C$) of $\sim 275$~K using an inflection point of \textit{M(T)} data as a criterion. This value will be determined more precisely below to be \textit{T}$_C$ = 274~K using easy axis \textit{M(H)} isotherms around Curie temperature.
 
\indent
FIG. \ref{Combinedgraph}(b) shows the anisotropic, field-dependent magnetization at $2$~K. The saturation magnetization ($M_{sat}$) at $2$~K is determined to be $2.40$~$\mu_{B}$/f.u., i.e. roughly half of bulk BCC Fe moment. The anisotropic \textit{M(H)} data at 2~K show [100] is the easy axis, [010] axis is a harder axis with an anisotropy field of $\approx 1$~T and [001] is the hardest axis of magnetization with an anisotropy field of $\approx 5$~T. A Sucksmith-Thompson  plot~\cite{Sucksmith362}, using \textit{M(H}) data along [001], is shown in FIG. \ref{Combinedgraph}(c). The inset to FIG. \ref{Combinedgraph}(c) shows data for \textit{H} along [010]. In a Sucksmith-Thompson plot, the Y-intercept of the linear fit of hard axis $\frac{H_{int}}{M}$ vs $M^2$ isotherm provides the magneto-crystalline anisotropy constant ($ intercept = \frac{2K_1}{{M_S}^2}$, $M_S$ being saturation magnetization at 2 K) of the material.  From these plots we determined $K_{010}$ = $0.23~MJm^{-3}$ and $K_{001}$ = $1.78~MJm^{-3}$ respectively. 
   
\begin{figure}[!h]
\begin{center}
\includegraphics[width=10cm]{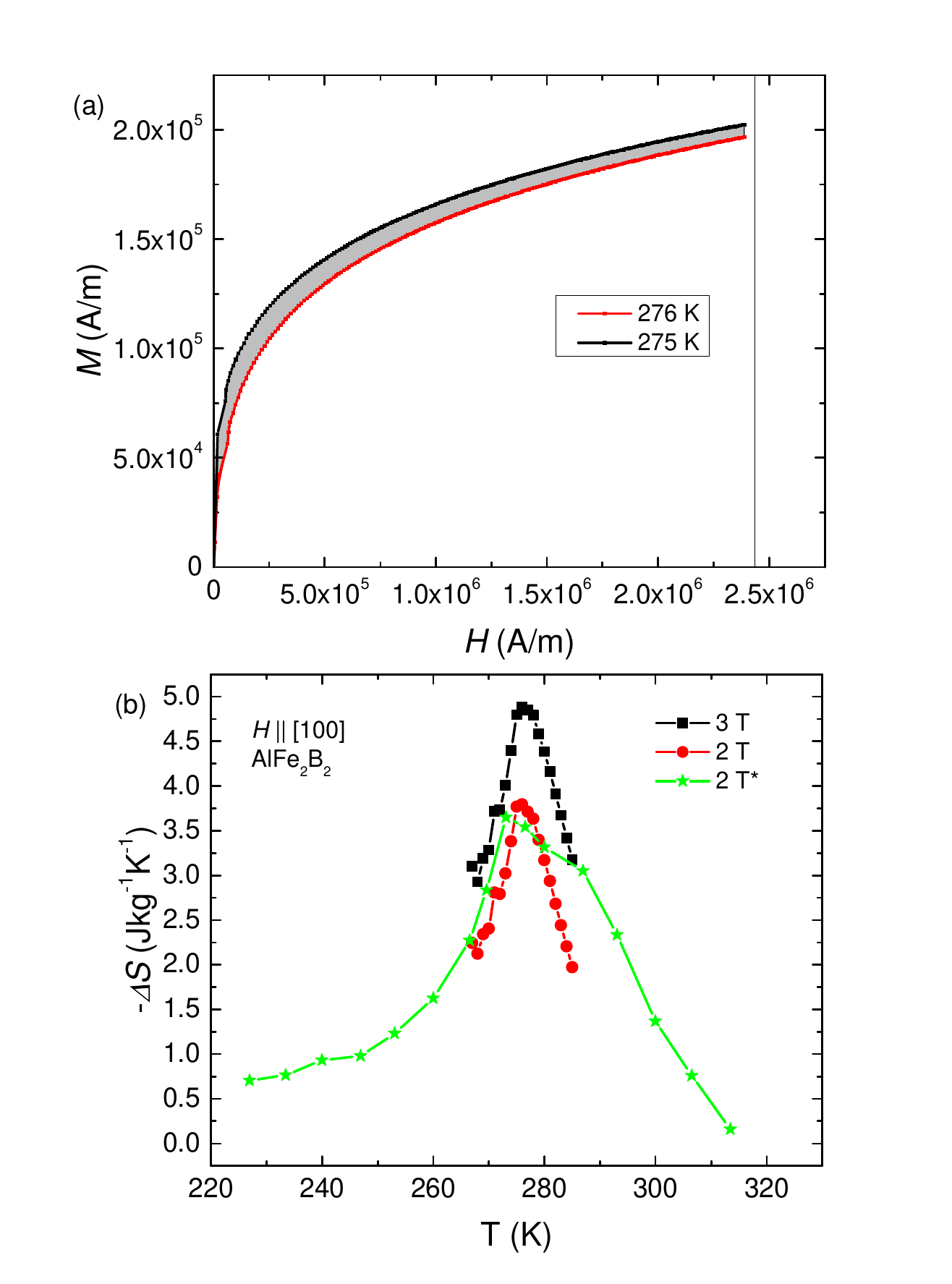}
\caption{Magnetocaloric effect in AlFe$_{2}$B$_{2}$ obtained using \textit{M(H)} isotherms along [100]. (a) showing the change in entropy ($\Delta S$) evaluation scheme at its highest value (b) change in entropy with $2$~T and $3$~T applied fields using easy axis [100] isotherms. For the sake of comparison, the 2 T$^*$  field data are taken from the reference [\onlinecite{BARUA2018}].}
\label{MCE-a-3T}
\end{center}
\end{figure}

Given that AlFe$_{2}$B$_{2}$ has $T_C\sim$ room temperature, and is formed from earth abundant elements, it is logical to examine it as a possible magnetocaloric material. The easy axis, [100], \textit{M(H)} isotherms around the Curie temperature (shown for the Arrott plot in FIG.~\ref{Combinedgraph}(d)) were used to estimate the magnetocaloric property for AlFe$_{2}$B$_{2}$ in-terms of entropy change using following equation \cite{Tanetal.AlFe2B2,HfZrMnPTej}:

\begin{equation}
\begin{split}
\Delta S(\frac{T_1 + T_2}{2},\Delta H)\approx \\
 &\dfrac{\mu_0}{T_2-T_1}\int^{H_f}_{H_i}M(T_2,H)-M(T_1,H)dH \\
 \label{eq:MaxwellMH}
\end{split}
\end{equation}

where $H_i$, $H_f$ are initial and final applied fields and $T_2-T_1$ is the change in temperature. For this formula to be valid, $T_2-T_1$  should be small. Here $T_2-T_1$ is taken to be $1$~K. The entropy change calculation scheme in one complete cycle of magnetization and demagnetization is estimated in terms of area between two consecutive isotherms between the given field limit as shown in FIG.~\ref{MCE-a-3T}(a). The measured entropy change as a function of temperature is presented in FIG.~\ref{MCE-a-3T}(b). The entropy change in $2$~T and $3$~T applied fields is maximum around $276$~K being $3.78$~Jkg$^{-1}$K$^{-1}$  and $4.87$~Jkg$^{-1}$K$^{-1}$respectively. The 2 T applied field entropy change data of this experiment agrees very well with reference [\onlinecite{BARUA2018}], shown as 2 T$^*$ data in FIG.~\ref{MCE-a-3T}(b). The entropy change values for our single crystalline samples are in close agreement with previously reported polycrystalline sample measured values as well \cite{Tanetal.AlFe2B2,QianhengDuMndopedAlFe2B2}.

Although, AlFe$_{2}$B$_{2}$ is a rare-earth free material, its magnetocaloric property is larger than  lighter rare-earth RT$_2$X$_2$ (R = rare earth T = transition metal, X = Si,Ge) compounds with  ThCr$_2$Si$_2$-type structure (space group I4/mmm) namely CeMn$_2$Ge$_2$($\sim1.8$~Jkg$^{-1}$K$^{-1}$)~\cite{MdDin2015}, PrMn$_2$Ge$_{0.8}$Si$_{1.2}$($\sim1.0$~Jkg$^{-1}$K$^{-1}$)~\cite{JLWang2009} and  Nd(Mn$_{1-x}$Fe$_x$)$_2$Ge$_2$($\sim1.0$~Jkg$^{-1}$K$^{-1}$)~\cite{CHEN201013}. The entropy change of AlFe$_2$B$_2$ is significantly smaller than Gd$_5$Si$_2$Ge$_2$($\sim 13$~Jkg$^{-1}$K$^{-1}$), it has comparable entropy change with elemental Gd($\sim 5.0$~Jkg$^{-1}$K$^{-1}$)~\cite{Pecharsky1997}. These results shows that AlFe$_2$B$_2$ has the potential to be used for magnetocaloric material considering the abundance of its constituents.

\begin{figure}
\begin{center}
\includegraphics[width=7cm]{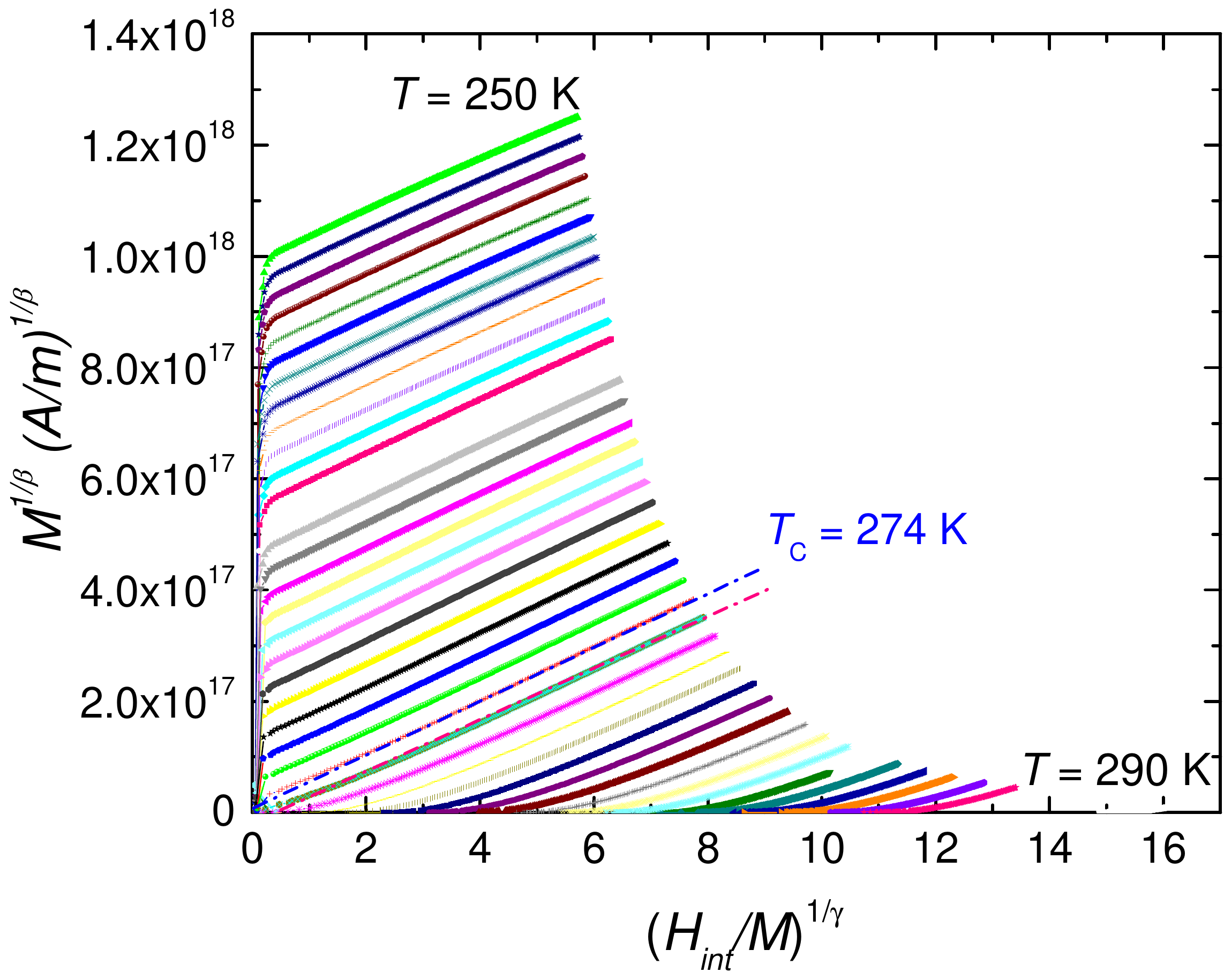}
\caption{Generalized Arrott plot of AlFe$_{2}$B$_{2}$ with magnetization data along [100] direction within temperature range of 250 - $290$~K at a step of $1$~K. The $\beta = 0.30 \pm 0.04$  and $\gamma= 1.180 \pm 0.005$  were determined from the Kouvel-Fisher method. The two dash-dot straight lines are drawn to visualize the intersection of the isotherms with the axes.}
\label{generalizedArrott}
\end{center}
\end{figure}

\begin{figure}[h]
  \includegraphics[width=7cm]{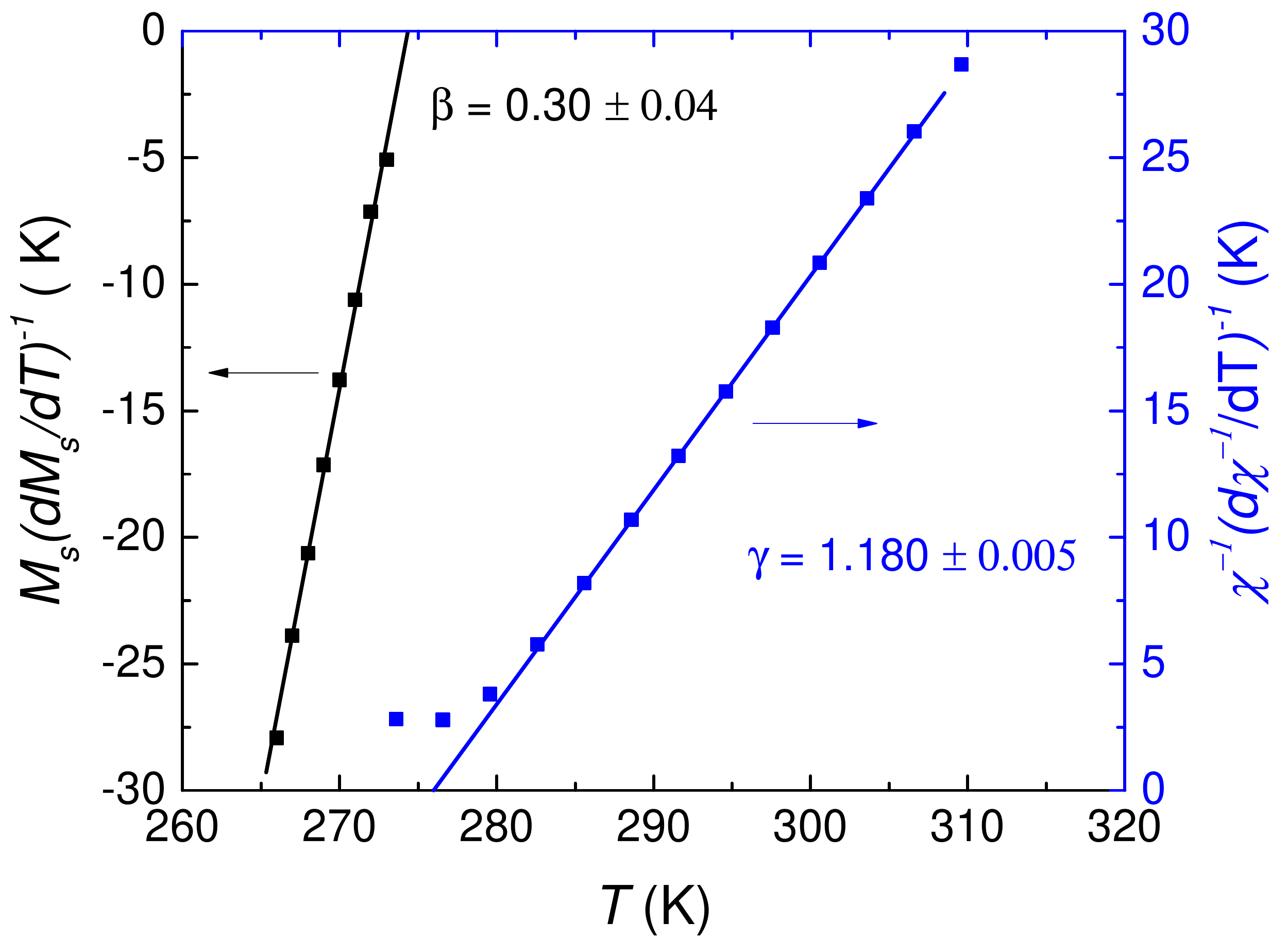}
  \caption{Determination of the critical exponents ($\beta$ and $\gamma$)using Kouvel-Fisher plots. See text for details.}
  \label{Beta-Gamma}
\end{figure}

\begin{figure}[h]
  \includegraphics[width=7cm]{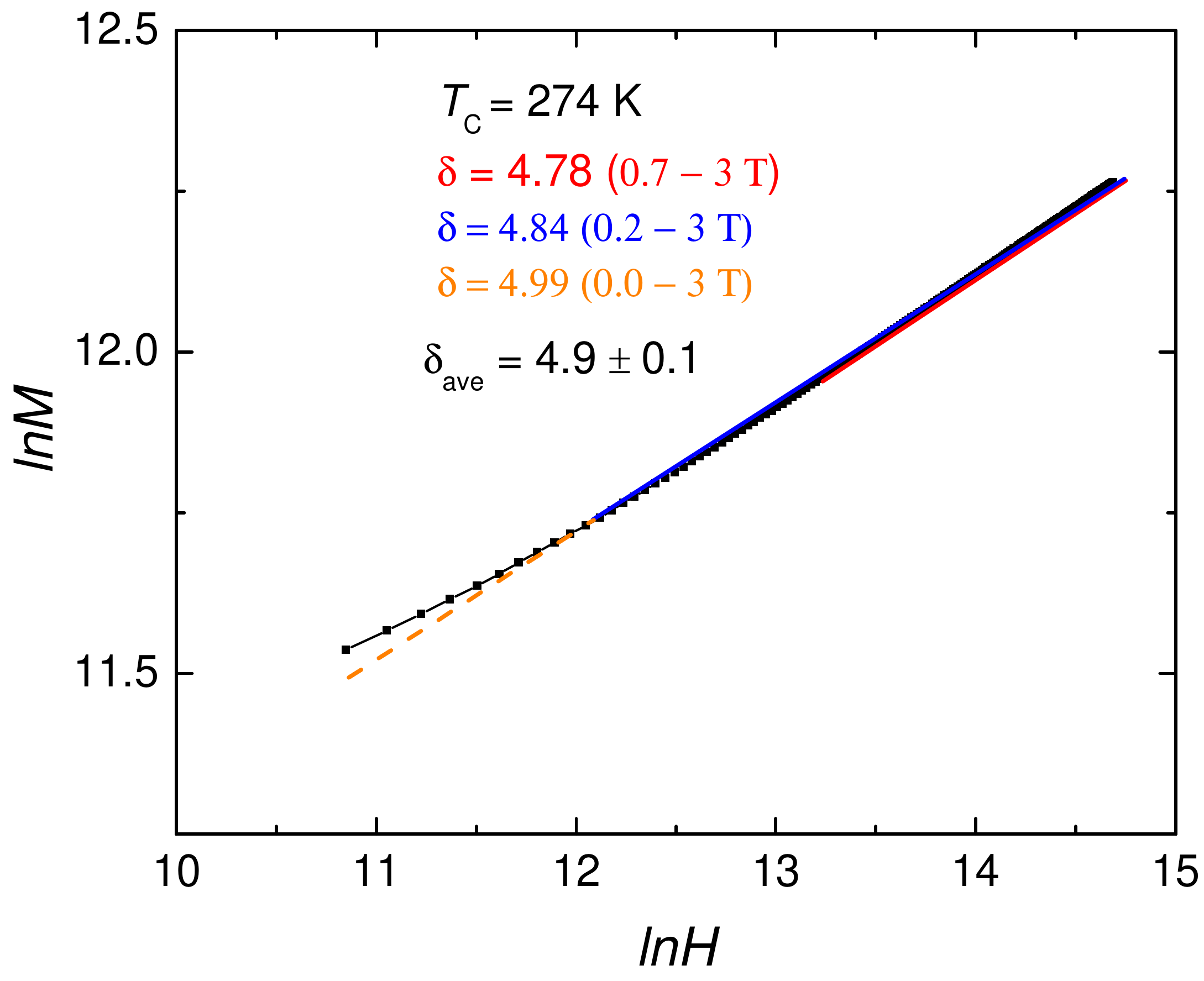}
  \caption{Determination of the critical exponent $\delta$ using Kouvel-Fisher plots using \textit{M(H)} isotherm at $T_C$ to check the consistency of $\beta$ and $\gamma$ via Widom scaling. The data used for determining the exponent $\delta$ are highlighted with the red curve in corresponding \textit{M(H)} isotherm. The data in the low field region slightly deviate from the linear behaviour in the logarithmic scale as shown in the inset. The range dependency of the value of  $\delta$ is illustrated with different colors tangents. The field range for the fitted data is indicated in the parenthesis along with the value of $\delta$.  See text for details.}
  \label{DeltaMH}
\end{figure}

\begin{figure}
\begin{center}
\includegraphics[width=7cm]{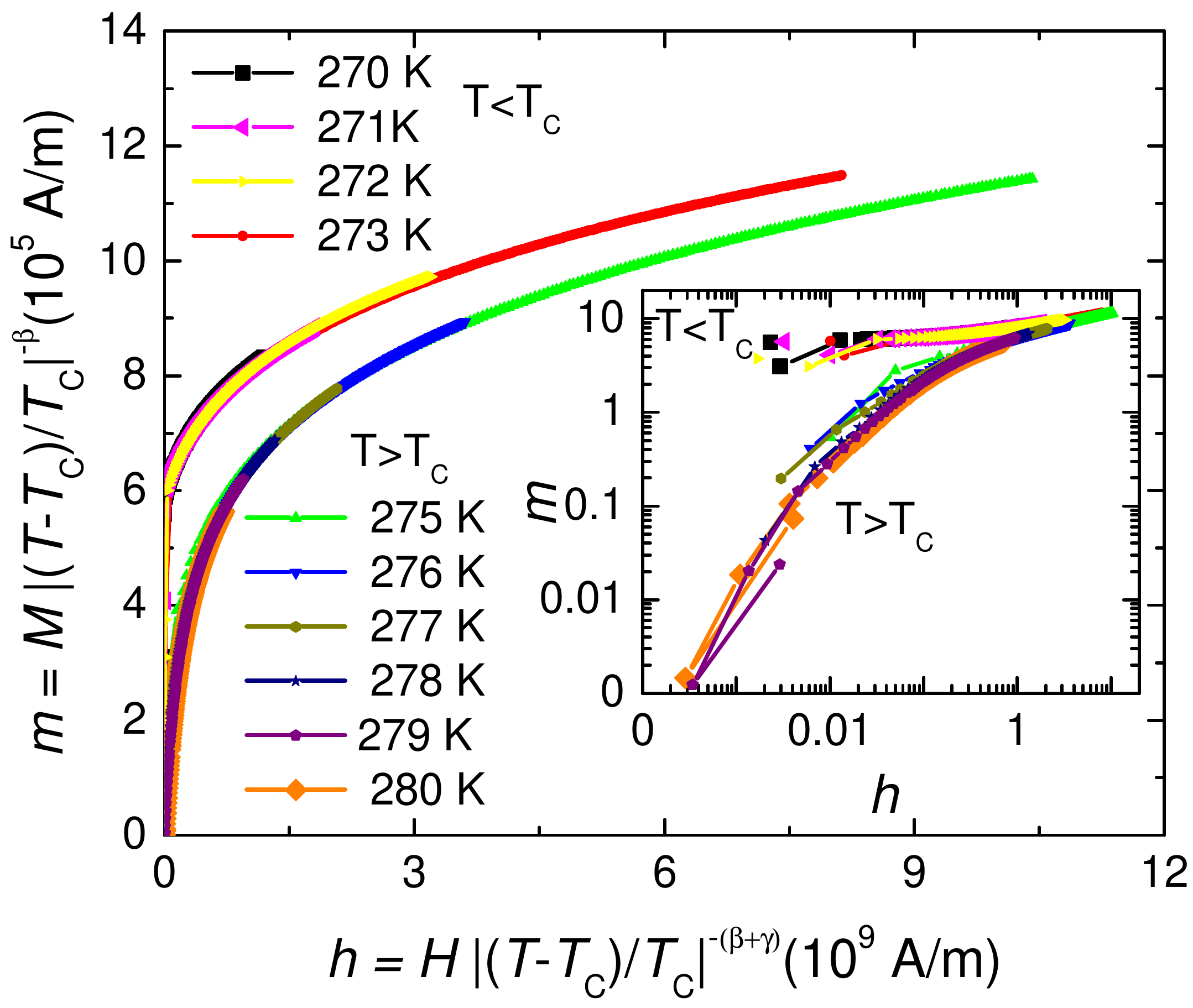}
\caption{Normalized isotherms to check the validity of  scaling hypothesis. The isotherms in between 270-$273$~K are converged to higher value (T$<$T$_C$) and isothems inbetween 275-$280$~K are converged to lower value(T$>$T$_C$). The inset shows the corresponding log-log plot clearly bifurcated in two branches in low field region.}
\label{Normalizedisotherms}
\end{center}
\end{figure}

\begin{figure}[!ht]
\begin{center}
\includegraphics[width=8cm]{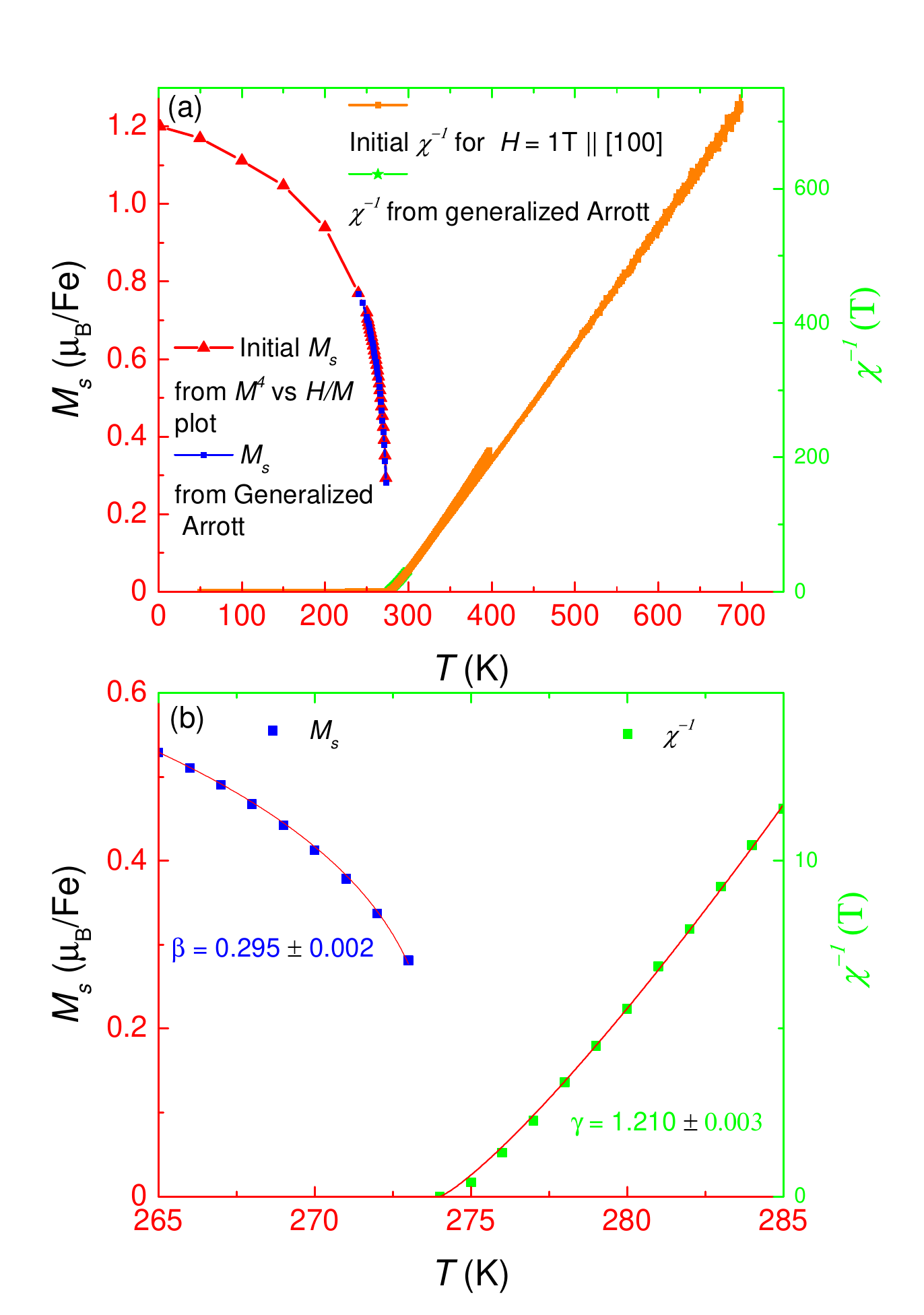}
\caption{Illustration of the consistency of the critical exponents $\beta$ and $\gamma$ used for generalizes Arrott plot (a) by reproducing the initial spontaneous magnetization $M_S$ and $\chi^{-1}(T)$ via Y and X-intercept of the generalized Arrott plot. (b) Fitting of the extracted data (squares) from generalized Arrott plot with corresponding power laws (red lines) in equation 6 and 7.}
\label{ExtractedMSHM}
\end{center}
\end{figure}

To precisely determine the Curie temperature, an Arrott plot was constructed using a wider range of \textit{M(H)} isotherms along the [100] direction (FIG.~\ref{Combinedgraph}(d)). In an Arrott plot $M^2$ is plotted as a function of $\frac{H_{int}}{M}$.\textit{ H$_{int}$}= \textit{H$_{app}$}-N*\textit{M} is internal field inside the sample after the demagnetization field is subtracted. In this case the experimental demagnetization factor along the easy axis of the sample was found to be almost negligible because of its thin, plate-like shape with the easy axis lying along the longest dimension of the sample. The detail of determination of the experimental demagnetization factors and their comparison with theoretical data is explained in the references~[\onlinecite{HfZrMnPTej}] and~[\onlinecite{Lamichhane2015}]. The Arrott plots have a positive slope indicating the transition is second order \cite{BANERJEE1964}. In the mean field approximation, in the limit of low fields, the Arrott isotherm corresponding to the Curie temperature is a straight line and passes through the origin. In FIG.~\ref{Combinedgraph}(d), the isotherm corresponding to $276$~K passes through the origin but it is not a perfectly straight line. This suggests that the magnetic interaction in AlFe$_{2}$B$_{2}$ does not obey the mean-field theory. In the mean-field theory, electron correlation and spin fluctuations are neglected,  but these can be significant around the transition temperature of an itinerant ferromagnet.

Since the Arrott plot data are not straight lines, a generalized Arrott plot is an alternative way to better confirm the Curie temperature. The  generalized Arrott plot derived from the equation of the state~\cite{Arrott1967} 

\begin{eqnarray}
({\frac{H_{int}}{M}})^{1/\gamma} = a \frac{T-T_{C}}{T} + b M^{1/\beta}
\end{eqnarray} 
 \noindent
is shown in FIG.~\ref{generalizedArrott}. The critical exponents $\beta$ and $\gamma$ used in equation of state are derived from the Kouvel-Fisher analysis\cite{KF1964,MohanKFLaSrMNO3}. To determine $\beta$, the equation used was:

\begin{eqnarray}
M_S[\frac{d}{dT}(M_S)]^{-1} = \frac{T-T_C}{\beta}
\end{eqnarray}

\noindent
where the slope is $\frac{1}{\beta}$. The value of the spontaneous magnetization around the transition temperature was extracted from the Y-intercept of the ${M_S}^4$ vs $\frac{H}{M}$~\cite{BinChenFe3GeTe2} exploiting their straight line nature with clear Y-intercept. The experimental value of $\beta$ was determined to be $0.30\pm0.04$ as shown in FIG.~\ref{Beta-Gamma}. The uncertainty in $\beta$ was determined with fitting error as $\Delta \beta$ = $\frac{\delta slope}{{slope}^2}$. 

\indent
Similarly, the value of critical exponent $\gamma$ was determined with the equation:
\begin{eqnarray}
{\chi}^{-1}[\frac{d}{dT}({\chi}^{-1})]^{-1} = \frac{T-T_C}{\gamma}
\end{eqnarray}
where the slope is $\frac{1}{\gamma}$ and $\chi^{-1}(T)$ is the initial high temperature inverse susceptibility near the transition temperature. 
The experimental value of the $\gamma$ was determined to be $1.180\pm 0.005$ as shown in FIG.~\ref{Beta-Gamma}.  


\indent
Finally the third critical exponent $\delta$ was determined using the equation:
\begin{eqnarray}
M\propto H^{1/\delta}
\end{eqnarray}
by plotting ln(\textit{M}) vs ln(\textit{H}) (FIG.~\ref{DeltaMH}) corresponding to Curie temperature $274$~K. The experimental value of the $\delta$ was determined  by fitting ln(\textit{M}) vs ln(\textit{H}) over different ranges of applied field \textit{H}. Taking the average of the range of the $\delta$ value as shown in FIG.~\ref{DeltaMH} we determine $\delta$ to be $4.9\pm 0.1$ which was closely reproduced ($4.93 \pm 0.03$) with Widom scaling theory $\delta = 1 + \frac{\gamma}{\beta}$.

\indent
 Additionally, the validity of Widom scaling theory demands that the magnetization data should follow the scaling equation of the state. The scaling laws for a second order magnetic phase transition relate the spontaneous magnetization \textit{M$_S$(T)} below \textit{T$_C$}, the inverse initial susceptibility ${\chi}^{-1}(T)$ above \textit{T$_C$}, and the magnetization at \textit{T$_C$} with corresponding critical amplitudes by the following power laws:
\begin{eqnarray}
M_S(T) = M_0(-\epsilon)^\beta, \epsilon < 0
\end{eqnarray}
 \begin{eqnarray}
{\chi}^{-1}(T) = \Gamma(\epsilon)^\gamma, \epsilon > 0
\end{eqnarray}
\begin{eqnarray}
M = XH^{\frac{1}{\delta}}
\end{eqnarray}
\noindent
Where M$_0$, $\Gamma$ and X are the critical amplitudes and $\epsilon = \frac{T-T_C}{T_C}$ is the reduced temperature \cite{ParamanikandBanarjee}.
The scaling hypothesis assumes the homogeneous order parameter which with scaling hypothesis can be expressed as 
\begin{eqnarray}
M(H,\epsilon) = {\epsilon}^{\beta}f_{\pm}(\frac{H}{{\epsilon}^{\beta + \gamma}})
\label{scalequ}
\end{eqnarray}
Where $f_{+}$(T$>T_C$) and $f_{-}$(T$<T_C$) are the regular functions. 
With new renormalised parameters, m = ${\epsilon}^{-\beta}M(H,\epsilon)$ and h = ${\epsilon}^{-(\beta+\gamma)}M(H,\epsilon)$ equation~\ref{scalequ} can be written as 
\begin{eqnarray}
m =f_{\pm}(h)
\end{eqnarray}  
Up to the linear order, the scaled  m vs h graph is plotted as shown in FIG~\ref{Normalizedisotherms}  along with an inset in log-log scale which clearly shows that all isotherms converge to the two curves one for T$>T_C$ and other for T$<T_C$. This graphically shows that all the critical exponents were properly renormalized.

Finally the consistency of the critical exponents $\beta$ and $\gamma$ is demonstrated (shown in FIG.~\ref{ExtractedMSHM}(a)) by reproducing the initial spontaneous magnetization $M_S$ and $\chi^{-1}(T)$ near the transition temperature using the Y and X-intercept of generalized Arott plots as shown in FIG.~\ref{generalizedArrott} which overlaps with $M_S$ obtained by ${M_S}^4$ vs $\frac{H}{M}$~[\onlinecite{BinChenFe3GeTe2}] and and initial inverse susceptibility $\chi^{-1}(T)$ with 1 T applied field. The extracted data well fit~\cite{ParamanikandBanarjee} with corresponding power laws in equation (6) and (7) as shown in FIG.~\ref{ExtractedMSHM}(b) giving $\beta = 0.295 \pm 0.002$ and $\gamma = 1.210 \pm 0.003$ which closely agree with previously obtained K-F values. 

To measure the effective moment ($M_{eff}$) of the Fe above the Curie temperature, a Curie-Weiss plot was prepared as shown in FIG.~\ref{ExtractedMSHM}. The effective moment of the Fe-ion above the Curie temperature was determined to be $2.15$~$\mu_{B}$. Since the effective moment above the Curie temperature is almost equal to BCC Fe ($2.2$~$\mu_{B}$) and the ordered moment at $2$~K is significantly smaller than Fe-ion ($1.2$~$\mu_{B}$/Fe) giving the Rhode-Wohlfarth ratio ($ \frac{M_{C}}{M_{sat}}$) nearly equal to 1.14, where $M_{C}(M_{C}+2)= {M_{eff}}^2$, this compound shows  signs of an itinerant nature in its magnetization~\cite{RWratio1963}. 

\indent

Itinerant magnetism, in general, can be tuned (meaning the size of magnetic moment and Curie temperature can be altered significantly and sometime even suppressed completely) with an external parameter like pressure or chemical doping. As a case study, we investigated the influence of the external pressure on the ferromagnetism of  AlFe$_{2}$B$_{2}$.

\begin{figure}
	\includegraphics[width=8.6cm]{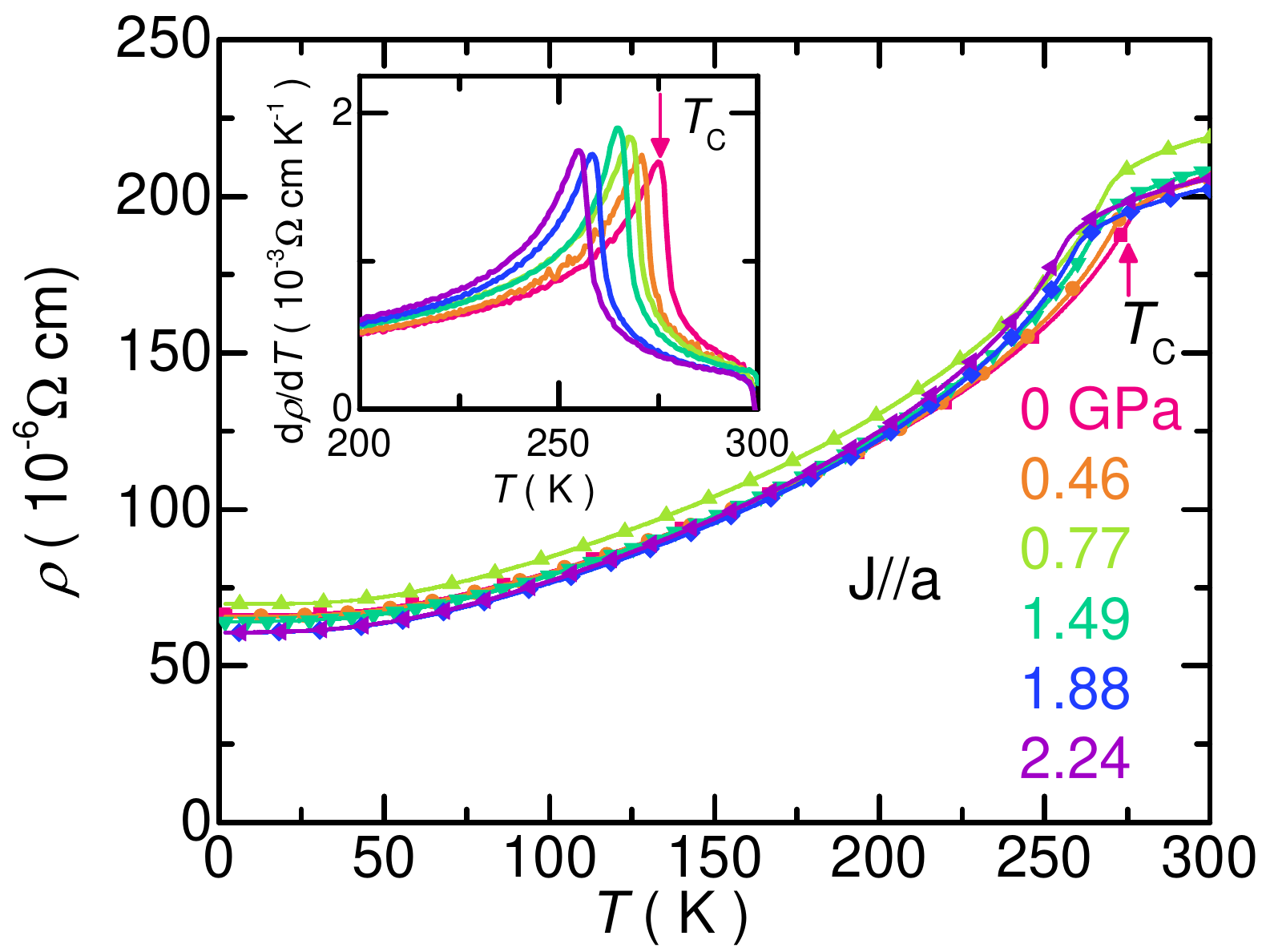}%
	\caption{Evolution of the single crystal AlFe$_{2}$B$_{2}$ resistivity with hydrostatic pressure up to $2.24$ GPa. Pressure values at $T_C$ were estimated from linear interpolation between the $P_{300\textrm{K}}$ and $P_{T\leq90\textrm{K}}$ values (see text). Current was applied along the crystallographic $a$-axis. Inset shows the evolution of temperature derivative $d\rho$/$dT$ with hydrostatic pressure. The peak positions in the derivative were identified as transition temperature $T_C$. Examples of $T_C$ are indicated by arrows in the figure. 
		\label{RT_pressure}}
\end{figure}

\begin{figure}
	\includegraphics[width=8.6cm]{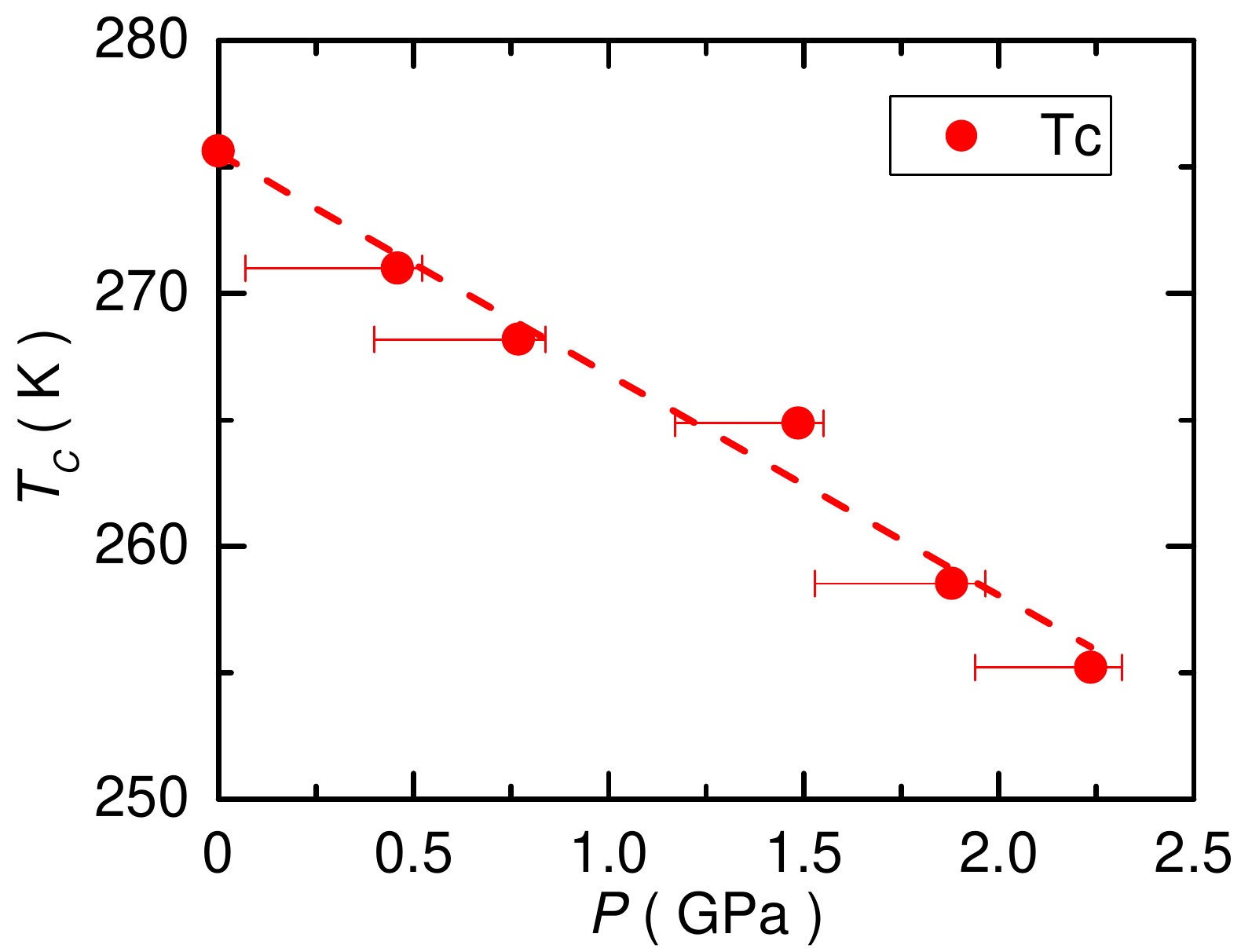}%
	\caption{Temperature - pressure phase diagram of AlFe$_2$B$_2$ as determined from resistivity measurement. Pressure values were estimated as being described in Fig.\,\ref{RT_pressure} and in the text. Error bars indicate the room temperature pressure $P_{300\textrm{K}}$ and low temperature pressure $P_{T\leq90\textrm{K}}$. As shown in the figure, in the pressure region of $0 - 2.24$ GPa the ferromagnetic transition temperature $T_C$ is suppressed upon increasing pressure, with suppressing rate around $-8.9$ K/GPa.}
\label{TP_phasediagram}
\end{figure}

\indent
Figure\,\ref{RT_pressure} shows the pressure dependent resistivity of single crystalline AlFe$_{2}$B$_{2}$  with current applied along the crystallographic $a$-axis. It shows metallic behaviour with a residual resistivity of 60 $\mu \Omega$ cm.  The ambient pressure temperature dependent resistivity of  AlFe$_{2}$B$_{2}$ shows a kink around $275$~K, indicating a loss of spin disorder scattering associated with the onset of ferromagnetic order. As pressure is increased to 2.24 GPa the temperature of this kink is steadily reduced. To determine the transition temperature $T_C$, the maximum in the temperature derivative $d\rho$/$dT$ is used, as shown in the inset of FIG.~\ref{RT_pressure}. The pressure dependence of $T_C$, the temperature - pressure phase diagram of AlFe$_2$B$_2$ is presented in FIG.~\ref{TP_phasediagram}. The transition temperature, $T_C$, is suppressed from $275$ K to $255$ K when pressure is increased from 0 to $2.24$ GPa, giving a suppression rate of -8.9 K/GPa. Interestingly, Curie temperature suppression rate of AlFe$_{2}$B$_{2}$ is found to be comparable to the model itinerant magnetic materials like helimagnetic MnSi ($\sim -15 $~K/GPa)~\cite{Pfleiderer2004}, and weak ferromagnets ZrZn$_2$ ($\sim -13$~K/GPa)~\cite{Uhlarz2004} and Ni$_3$Al ($\sim -4$~K/GPa)~\cite{Niklowitz2005}. A linear fitting of the data as shown in FIG.~\ref{TP_phasediagram} indicates that to completely suppress the $T_C$ around 31 GPa would be required. Usually such linear extrapolation provide an upper estimate of the critical pressure. 

\begin{figure}[!ht]
\centering
\includegraphics[width=\columnwidth]{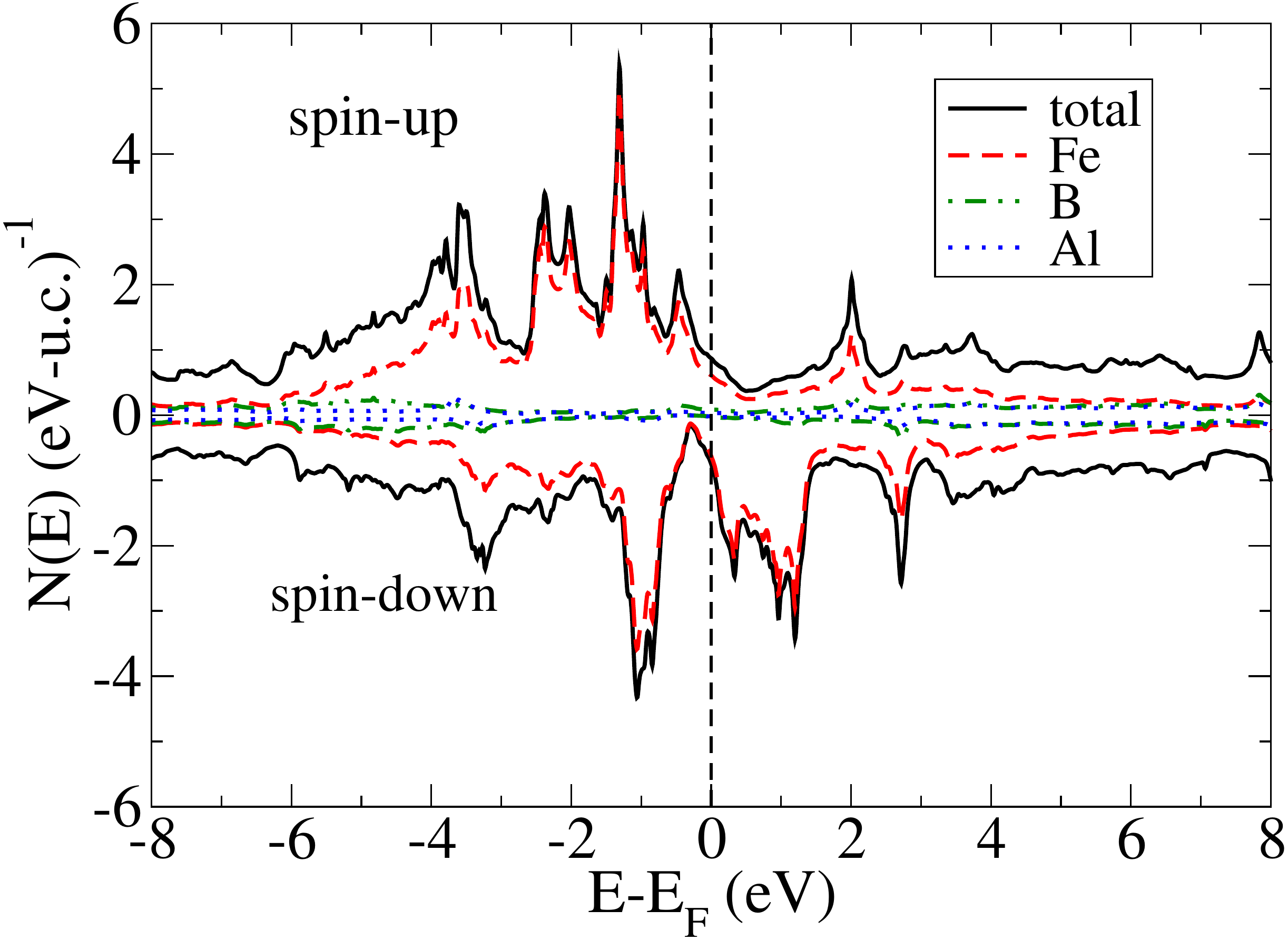}
\caption{The calculated density-of-states of AlFe$_2$B$_2$.}
\label{fig:DOS}
\end{figure}

\section*{First principles calculations}
Theoretical calculations for AlFe$_2$B$_2$ were performed using the all electron density functional theory code WIEN2K~\cite{wien2k,singh_lo,singh_book}. The generalized gradient approximation according to Perdew, Burke, and Ernzerhof (PBE)~\cite{perdew1996generalized} was used in our calculations. The sphere radii (RMT) were set to 2.21, 2.17, and 1.53 Bohr for Fe, Al, and B, respectively. RK$_{max}$ which defines the product of the smallest sphere radius and the largest plane wave vector was set to 7.0. All calculations were performed with the experimental lattice parameters as reported in reference~[\onlinecite{cedervall2016magnetic}] (which are consistent with our results) and all internal coordinates were relaxed until internal forces on atoms were less than 1 mRyd/Bohr-radius. All the calculations were performed in the collinear spin alignment. The magnetic anisotropy energy (MAE) was obtained by calculating the total energies of the system with spin-orbit coupling (SOC) with the magnetic moment along the three principal crystallographic axes. For these MAE calculations the $k$-point convergence was carefully checked, and the calculations reported here were performed with 120,000 $k$-points in the full Brillouin zone.

Similar to the experimental observation, AlFe$_2$B$_2$ is calculated to have ferromagnetic behaviour, with a saturation magnetic moment (we do not include the small Fe orbital moment) of 1.36 $\mu_B$/Fe. This is in reasonable agreement with the experimentally measured value of 
 1.21$\mu$B/Fe. Interestingly this calculated magnetic moment on Fe, is significantly lower than the moment on Fe in BCC Fe (2.2 $\mu_B$/Fe ) further suggesting a degree of itinerant behaviour. The calculated density of states is shown in FIG.~\ref{fig:DOS}. As expected for a Fe-based ferromagnet, the electronic structure in the vicinity of the Fermi level is dominated by Fe $d$ orbitals and we observe a substantial exchange splitting of 2-3 eV. 

For an orthorhombic crystal structure, the magnetic anisotropy energy is described by total energy calculations for the magnetic moments along each of the three principal axis~\cite{palmer1963magnetocrystalline}. 
For AlFe$_2$B$_2$ we find, the [100] and [010] axes to be the ``easy" directions, separated by just 0.016 meV per Fe, with [100] being the easiest axis. The [001] direction is the ``hard" direction, which lies 0.213 meV per Fe above the [100] axis.  As in our previous work on HfMnP~\cite{HfZrMnPTej}, this value is much larger than the 0.06 meV value for hcp Co and likely results from a combination of the orthorhombic crystal structure and the structural complexity associated with a ternary compound. The 0.213 meV energy difference on a volumetric basis corresponds to a anisotropy constant  K$_ 1$ as 1.48 MJ/m$^3$. (Note that we use the convention of the previous work and simply {\it define} K$_{1}$ for an orthorhombic system as the energy difference between the hardest and easiest directions.) This magnetic anisotropy constant describes the energy cost associated with the changing the orientation of the magnetic moments under the application of a magnetic field, and is an essential component for permanent magnets. It is noteworthy that this anisotropy is comparable to the value of 2 MJ/m$^3$ proposed by Coey for an efficient permanent magnet~\cite{coey2011hard}, despite containing no heavy elements, using the approximation that $H_a \approx 2 \mu_0 K_1/M_s$, with K as 1.48 MJ/m$^3$ and M$_s$ as 0.68 T, yields an anisotropy field of 5.4 T, which is in excellent agreement with the experimentally measured value of 5 T and $K_{001} \approx 1.8~MJ/m^{3}$.

\section{Conclusions}
Single crystalline  AlFe$_{2}$B$_{2}$ was grown by self-flux-growth technique and structural, magnetic and transport properties were studied. AlFe$_{2}$B$_{2}$ is an orthorhombic, metallic ferromagnet with promising magnetocaloric behaviour. The Curie temperature of AlFe$_{2}$B$_{2}$ was determined to be $274$~K using the generalized Arrott plot method along with estimation of critical exponents using  Kouvel-Fisher analysis. The ordered magnetic moment ($M_{sat}$) at $2$~K is $1.20\mu_{B}$/Fe at $2$~K which is much less than paramagnetic Fe-ion moment at high temperature ($2.15\mu_{B}$/Fe) indicating itinerant magnetism. The magnetization in AlFe$_{2}$B$_{2}$ responds to the hydrostatic pressure with $\frac{dT_C}{dP} \sim -8.9$~K/GPa. A linear extrapolation of this $T_C(P)$ trend leads to an upper estimate of  $\sim30$ GPa required to fully supress the transition. The saturation magnetization and anisotropic magnetic field predicted by first principle calculations are in close agreement with the experimental results. The magneto-crystalline anisotropy fields were determined to be $1$~T along [010] and $5$~T along [001] direction w. r. to easy axis [100]. The magneto-crystalline anisotropy constants at 2 K are determined to be $K_{010} \approx 0.23~MJ/m^3$ and $K_{001} \approx 1.8~MJ/m^3$.   
 
  
\section{Acknowledgement}
We would like to thank Drs. W. R. McCallum and L. H. Lewis for drawing attention to this compound and Dr. A. Palasyuk for useful discussions. This research was supported by the Critical Materials Institute, an Energy Innovation Hub funded by the U.S. Department of Energy, Office of Energy Efficiency and Renewable Energy, Advanced Manufacturing Office. This work was also supported by the office of Basic Energy Sciences, Materials Sciences Division, U.S. DOE. Li Xiang was supported by W. M. Keck Foundation. This work was performed at the Ames Laboratory, operated for DOE by Iowa State University under Contract No. DE-AC02-07CH11358.\\*

\bibliography{TejResearch-AlFe2B2}

\begin{thebibliography}{43}%
\makeatletter
\providecommand \@ifxundefined [1]{%
 \@ifx{#1\undefined}
}%
\providecommand \@ifnum [1]{%
 \ifnum #1\expandafter \@firstoftwo
 \else \expandafter \@secondoftwo
 \fi
}%
\providecommand \@ifx [1]{%
 \ifx #1\expandafter \@firstoftwo
 \else \expandafter \@secondoftwo
 \fi
}%
\providecommand \natexlab [1]{#1}%
\providecommand \enquote  [1]{``#1''}%
\providecommand \bibnamefont  [1]{#1}%
\providecommand \bibfnamefont [1]{#1}%
\providecommand \citenamefont [1]{#1}%
\providecommand \href@noop [0]{\@secondoftwo}%
\providecommand \href [0]{\begingroup \@sanitize@url \@href}%
\providecommand \@href[1]{\@@startlink{#1}\@@href}%
\providecommand \@@href[1]{\endgroup#1\@@endlink}%
\providecommand \@sanitize@url [0]{\catcode `\\12\catcode `\$12\catcode
  `\&12\catcode `\#12\catcode `\^12\catcode `\_12\catcode `\%12\relax}%
\providecommand \@@startlink[1]{}%
\providecommand \@@endlink[0]{}%
\providecommand \url  [0]{\begingroup\@sanitize@url \@url }%
\providecommand \@url [1]{\endgroup\@href {#1}{\urlprefix }}%
\providecommand \urlprefix  [0]{URL }%
\providecommand \Eprint [0]{\href }%
\providecommand \doibase [0]{http://dx.doi.org/}%
\providecommand \selectlanguage [0]{\@gobble}%
\providecommand \bibinfo  [0]{\@secondoftwo}%
\providecommand \bibfield  [0]{\@secondoftwo}%
\providecommand \translation [1]{[#1]}%
\providecommand \BibitemOpen [0]{}%
\providecommand \bibitemStop [0]{}%
\providecommand \bibitemNoStop [0]{.\EOS\space}%
\providecommand \EOS [0]{\spacefactor3000\relax}%
\providecommand \BibitemShut  [1]{\csname bibitem#1\endcsname}%
\let\auto@bib@innerbib\@empty
\bibitem [{\citenamefont {Tan}\ \emph {et~al.}(2013)\citenamefont {Tan},
  \citenamefont {Chai}, \citenamefont {Thompson},\ and\ \citenamefont
  {Michael}}]{Tanetal.AlFe2B2}%
  \BibitemOpen
  \bibfield  {author} {\bibinfo {author} {\bibfnamefont {Xiaoyan}\ \bibnamefont
  {Tan}}, \bibinfo {author} {\bibfnamefont {Ping}\ \bibnamefont {Chai}},
  \bibinfo {author} {\bibfnamefont {Corey~M.}\ \bibnamefont {Thompson}}, \ and\
  \bibinfo {author} {\bibfnamefont {Shatruk}\ \bibnamefont {Michael}},\
  }\bibfield  {title} {\enquote {\bibinfo {title} {{Magnetocaloric Effect in
  AlFe$_2$B$_2$: Toward Magnetic Refrigerants from Earth-Abundant Elements}},}\
  }\href {\doibase 10.1021/ja404107p} {\bibfield  {journal} {\bibinfo
  {journal} {J. Am. Chem. Soc.}\ }\textbf {\bibinfo {volume} {135}},\ \bibinfo
  {pages} {9553--9557} (\bibinfo {year} {2013})},\ \Eprint
  {http://arxiv.org/abs/http://dx.doi.org/10.1021/ja404107p}
  {http://dx.doi.org/10.1021/ja404107p} \BibitemShut {NoStop}%
\bibitem [{\citenamefont {Cedervall}\ \emph
  {et~al.}(2016{\natexlab{a}})\citenamefont {Cedervall}, \citenamefont
  {H{\"a}ggstr{\"o}m}, \citenamefont {Ericsson},\ and\ \citenamefont
  {Sahlberg}}]{Cedervall2016}%
  \BibitemOpen
  \bibfield  {author} {\bibinfo {author} {\bibfnamefont {Johan}\ \bibnamefont
  {Cedervall}}, \bibinfo {author} {\bibfnamefont {Lennart}\ \bibnamefont
  {H{\"a}ggstr{\"o}m}}, \bibinfo {author} {\bibfnamefont {Tore}\ \bibnamefont
  {Ericsson}}, \ and\ \bibinfo {author} {\bibfnamefont {Martin}\ \bibnamefont
  {Sahlberg}},\ }\bibfield  {title} {\enquote {\bibinfo {title} {{M{\"o}ssbauer
  study of the magnetocaloric compound AlFe$_2$B$_2$}},}\ }\href {\doibase
  doi.org/10.1007/s1075} {\bibfield  {journal} {\bibinfo  {journal} {Hyperfine
  Interactions}\ }\textbf {\bibinfo {volume} {237}},\ \bibinfo {pages} {47}
  (\bibinfo {year} {2016}{\natexlab{a}})}\BibitemShut {NoStop}%
\bibitem [{\citenamefont {ElMassalami}\ \emph {et~al.}(2011)\citenamefont
  {ElMassalami}, \citenamefont {da~S.~Oliveira},\ and\ \citenamefont
  {Takeya}}]{ElMassalamiAlFe2B2}%
  \BibitemOpen
  \bibfield  {author} {\bibinfo {author} {\bibfnamefont {M.}~\bibnamefont
  {ElMassalami}}, \bibinfo {author} {\bibfnamefont {D.}~\bibnamefont
  {da~S.~Oliveira}}, \ and\ \bibinfo {author} {\bibfnamefont {H.}~\bibnamefont
  {Takeya}},\ }\bibfield  {title} {\enquote {\bibinfo {title} {{On the
  ferromagnetism of AlFe$_2$B$_2$ }},}\ }\href {\doibase
  http://doi.org/10.1016/j.jmmm.2011.03.008} {\bibfield  {journal} {\bibinfo
  {journal} {J. Magn. Magn. Mater.}\ }\textbf {\bibinfo {volume} {323}},\
  \bibinfo {pages} {2133 -- 2136} (\bibinfo {year} {2011})}\BibitemShut
  {NoStop}%
\bibitem [{\citenamefont {Jeitschko}(1969)}]{WJeitschkoFe2AlB2}%
  \BibitemOpen
  \bibfield  {author} {\bibinfo {author} {\bibfnamefont {Wolfgang}\
  \bibnamefont {Jeitschko}},\ }\bibfield  {title} {\enquote {\bibinfo {title}
  {{The crystal structure of Fe$_2$AlB$_2$.}}}\ }\href@noop {} {\bibfield
  {journal} {\bibinfo  {journal} {Acta Cryst.}\ }\textbf {\bibinfo {volume}
  {B}},\ \bibinfo {pages} {163--165} (\bibinfo {year} {1969})}\BibitemShut
  {NoStop}%
\bibitem [{\citenamefont {Kuz'ma}\ and\ \citenamefont
  {Chaban}(1969)}]{Kuzma1969}%
  \BibitemOpen
  \bibfield  {author} {\bibinfo {author} {\bibfnamefont {Yu.~B.}\ \bibnamefont
  {Kuz'ma}}\ and\ \bibinfo {author} {\bibfnamefont {N.F.}\ \bibnamefont
  {Chaban}},\ }\bibfield  {title} {\enquote {\bibinfo {title} {{Crystal
  Structure of the Compound Fe$_{2}$AlB$_{2}$}},}\ }\href@noop {} {\bibfield
  {journal} {\bibinfo  {journal} {Inorg. Mater.}\ }\textbf {\bibinfo {volume}
  {5}},\ \bibinfo {pages} {321--322} (\bibinfo {year} {1969})}\BibitemShut
  {NoStop}%
\bibitem [{\citenamefont {Du}\ \emph {et~al.}(2015)\citenamefont {Du},
  \citenamefont {Chen}, \citenamefont {Yang}, \citenamefont {Wei},
  \citenamefont {Hua}, \citenamefont {Du}, \citenamefont {Wang}, \citenamefont
  {Liu}, \citenamefont {Han}, \citenamefont {Zhang},\ and\ \citenamefont
  {Yang}}]{QianhengDuMndopedAlFe2B2}%
  \BibitemOpen
  \bibfield  {author} {\bibinfo {author} {\bibfnamefont {Qianheng}\
  \bibnamefont {Du}}, \bibinfo {author} {\bibfnamefont {Guofu}\ \bibnamefont
  {Chen}}, \bibinfo {author} {\bibfnamefont {Wenyun}\ \bibnamefont {Yang}},
  \bibinfo {author} {\bibfnamefont {Jianzhong}\ \bibnamefont {Wei}}, \bibinfo
  {author} {\bibfnamefont {Muxin}\ \bibnamefont {Hua}}, \bibinfo {author}
  {\bibfnamefont {Honglin}\ \bibnamefont {Du}}, \bibinfo {author}
  {\bibfnamefont {Changsheng}\ \bibnamefont {Wang}}, \bibinfo {author}
  {\bibfnamefont {Shunquan}\ \bibnamefont {Liu}}, \bibinfo {author}
  {\bibfnamefont {Jingzhi}\ \bibnamefont {Han}}, \bibinfo {author}
  {\bibfnamefont {Yan}\ \bibnamefont {Zhang}}, \ and\ \bibinfo {author}
  {\bibfnamefont {Jinbo}\ \bibnamefont {Yang}},\ }\bibfield  {title} {\enquote
  {\bibinfo {title} {{Magnetic frustration and magnetocaloric effect in
  AlFe$_{2 - x}$ Mn$_x$ B$_2$ ( x = 0 - 0.5) ribbons}},}\ }\href
  {http://stacks.iop.org/0022-3727/48/i=33/a=335001} {\bibfield  {journal}
  {\bibinfo  {journal} {Journal of Physics D: Applied Physics}\ }\textbf
  {\bibinfo {volume} {48}},\ \bibinfo {pages} {335001} (\bibinfo {year}
  {2015})}\BibitemShut {NoStop}%
\bibitem [{\citenamefont {Hirt}\ \emph {et~al.}(2016)\citenamefont {Hirt},
  \citenamefont {Yuan}, \citenamefont {Mozharivskyj},\ and\ \citenamefont
  {Hillebrecht}}]{HirtAlFe2B2Co}%
  \BibitemOpen
  \bibfield  {author} {\bibinfo {author} {\bibfnamefont {Sarah}\ \bibnamefont
  {Hirt}}, \bibinfo {author} {\bibfnamefont {Fang}\ \bibnamefont {Yuan}},
  \bibinfo {author} {\bibfnamefont {Yurij}\ \bibnamefont {Mozharivskyj}}, \
  and\ \bibinfo {author} {\bibfnamefont {Harald}\ \bibnamefont {Hillebrecht}},\
  }\bibfield  {title} {\enquote {\bibinfo {title} {{AlFe$_{2 - x}$Co$_x$B$_2$
  (x = 0 - 0.30): T$_C$ Tuning through Co Substitution for a Promising
  Magnetocaloric Material Realized by Spark Plasma Sintering}},}\ }\href
  {\doibase 10.1021/acs.inorgchem.6b01467} {\bibfield  {journal} {\bibinfo
  {journal} {Inorg. Chem.}\ }\textbf {\bibinfo {volume} {55}},\ \bibinfo
  {pages} {9677--9684} (\bibinfo {year} {2016})},\ \bibinfo {note} {pMID:
  27622951},\ \Eprint
  {http://arxiv.org/abs/http://dx.doi.org/10.1021/acs.inorgchem.6b01467}
  {http://dx.doi.org/10.1021/acs.inorgchem.6b01467} \BibitemShut {NoStop}%
\bibitem [{\citenamefont {Barua}\ \emph {et~al.}(2018)\citenamefont {Barua},
  \citenamefont {Lejeune}, \citenamefont {Ke}, \citenamefont {Hadjipanayis},
  \citenamefont {Levin}, \citenamefont {McCallum}, \citenamefont {Kramer},\
  and\ \citenamefont {Lewis}}]{BARUA2018}%
  \BibitemOpen
  \bibfield  {author} {\bibinfo {author} {\bibfnamefont {R.}~\bibnamefont
  {Barua}}, \bibinfo {author} {\bibfnamefont {B.T.}\ \bibnamefont {Lejeune}},
  \bibinfo {author} {\bibfnamefont {L.}~\bibnamefont {Ke}}, \bibinfo {author}
  {\bibfnamefont {G.}~\bibnamefont {Hadjipanayis}}, \bibinfo {author}
  {\bibfnamefont {E.M.}\ \bibnamefont {Levin}}, \bibinfo {author}
  {\bibfnamefont {R.W.}\ \bibnamefont {McCallum}}, \bibinfo {author}
  {\bibfnamefont {M.J.}\ \bibnamefont {Kramer}}, \ and\ \bibinfo {author}
  {\bibfnamefont {L.H.}\ \bibnamefont {Lewis}},\ }\bibfield  {title} {\enquote
  {\bibinfo {title} {{Anisotropic magnetocaloric response in AlFe$_2$B$_2$}},}\
  }\href {\doibase https://doi.org/10.1016/j.jallcom.2018.02.205} {\bibfield
  {journal} {\bibinfo  {journal} {J. Alloys Compd.}\ }\textbf {\bibinfo
  {volume} {745}},\ \bibinfo {pages} {505 -- 512} (\bibinfo {year}
  {2018})}\BibitemShut {NoStop}%
\bibitem [{\citenamefont {Levin}\ \emph {et~al.}(2018)\citenamefont {Levin},
  \citenamefont {Jensen}, \citenamefont {Barua}, \citenamefont {Lejeune},
  \citenamefont {Howard}, \citenamefont {McCallum}, \citenamefont {Kramer},\
  and\ \citenamefont {Lewis}}]{Levin2018}%
  \BibitemOpen
  \bibfield  {author} {\bibinfo {author} {\bibfnamefont {E.~M.}\ \bibnamefont
  {Levin}}, \bibinfo {author} {\bibfnamefont {B.~A.}\ \bibnamefont {Jensen}},
  \bibinfo {author} {\bibfnamefont {R.}~\bibnamefont {Barua}}, \bibinfo
  {author} {\bibfnamefont {B.}~\bibnamefont {Lejeune}}, \bibinfo {author}
  {\bibfnamefont {A.}~\bibnamefont {Howard}}, \bibinfo {author} {\bibfnamefont
  {R.~W.}\ \bibnamefont {McCallum}}, \bibinfo {author} {\bibfnamefont {M.~J.}\
  \bibnamefont {Kramer}}, \ and\ \bibinfo {author} {\bibfnamefont {L.~H.}\
  \bibnamefont {Lewis}},\ }\bibfield  {title} {\enquote {\bibinfo {title}
  {{Effects of Al content and annealing on the phases formation, lattice
  parameters, and magnetization of A l$_ x $Fe$_2$ B$_2$( x = 1.0 , 1.1 , 1.2 )
  alloys}},}\ }\href {\doibase 10.1103/PhysRevMaterials.2.034403} {\bibfield
  {journal} {\bibinfo  {journal} {Phys. Rev. Materials}\ }\textbf {\bibinfo
  {volume} {2}},\ \bibinfo {pages} {034403} (\bibinfo {year}
  {2018})}\BibitemShut {NoStop}%
\bibitem [{\citenamefont {Canfield}\ and\ \citenamefont
  {Fisk}(1992)}]{CanfieldFisk}%
  \BibitemOpen
  \bibfield  {author} {\bibinfo {author} {\bibfnamefont {P.~C.}\ \bibnamefont
  {Canfield}}\ and\ \bibinfo {author} {\bibfnamefont {Z.}~\bibnamefont
  {Fisk}},\ }\bibfield  {title} {\enquote {\bibinfo {title} {{Growth of single
  crystals from metallic fluxes}},}\ }\href {\doibase
  10.1080/13642819208215073} {\bibfield  {journal} {\bibinfo  {journal}
  {{Philos. Mag.}}\ }\textbf {\bibinfo {volume} {65}},\ \bibinfo {pages}
  {1117--1123} (\bibinfo {year} {1992})},\ \Eprint
  {http://arxiv.org/abs/http://dx.doi.org/10.1080/13642819208215073}
  {http://dx.doi.org/10.1080/13642819208215073} \BibitemShut {NoStop}%
\bibitem [{\citenamefont {Canfield}\ \emph {et~al.}(2016)\citenamefont
  {Canfield}, \citenamefont {Kong}, \citenamefont {Kaluarachchi},\ and\
  \citenamefont {Jo}}]{Canfieldfrittedcrucible}%
  \BibitemOpen
  \bibfield  {author} {\bibinfo {author} {\bibfnamefont {Paul~C.}\ \bibnamefont
  {Canfield}}, \bibinfo {author} {\bibfnamefont {Tai}\ \bibnamefont {Kong}},
  \bibinfo {author} {\bibfnamefont {Udhara~S.}\ \bibnamefont {Kaluarachchi}}, \
  and\ \bibinfo {author} {\bibfnamefont {Na~Hyun}\ \bibnamefont {Jo}},\
  }\bibfield  {title} {\enquote {\bibinfo {title} {Use of frit-disc crucibles
  for routine and exploratory solution growth of single crystalline samples},}\
  }\href {\doibase 10.1080/14786435.2015.1122248} {\bibfield  {journal}
  {\bibinfo  {journal} {Philos. Mag.}\ }\textbf {\bibinfo {volume} {96}},\
  \bibinfo {pages} {84--92} (\bibinfo {year} {2016})}\BibitemShut {NoStop}%
\bibitem [{\citenamefont {Bud'ko}\ \emph {et~al.}(1984)\citenamefont {Bud'ko},
  \citenamefont {Voronovskii}, \citenamefont {Gapotchenko},\ and\ \citenamefont
  {ltskevich}}]{Budko1984}%
  \BibitemOpen
  \bibfield  {author} {\bibinfo {author} {\bibfnamefont {S.L.}\ \bibnamefont
  {Bud'ko}}, \bibinfo {author} {\bibfnamefont {A.N.}\ \bibnamefont
  {Voronovskii}}, \bibinfo {author} {\bibfnamefont {A.G.}\ \bibnamefont
  {Gapotchenko}}, \ and\ \bibinfo {author} {\bibfnamefont {E.S.}\ \bibnamefont
  {ltskevich}},\ }\bibfield  {title} {\enquote {\bibinfo {title} {The fermi
  surface of cadmium at an electron-topological phase transition under
  pressure},}\ }\href@noop {} {\bibfield  {journal} {\bibinfo  {journal} {J.
  Exp. Theor. Phys.}\ }\textbf {\bibinfo {volume} {59}},\ \bibinfo {pages}
  {454} (\bibinfo {year} {1984})}\BibitemShut {NoStop}%
\bibitem [{\citenamefont {Thompson}(1984)}]{Thompson1984}%
  \BibitemOpen
  \bibfield  {author} {\bibinfo {author} {\bibfnamefont {J.~D.}\ \bibnamefont
  {Thompson}},\ }\bibfield  {title} {\enquote {\bibinfo {title}
  {{Low-temperature pressure variations in a self-clamping pressure cell}},}\
  }\href {\doibase 10.1063/1.1137730} {\bibfield  {journal} {\bibinfo
  {journal} {Rev. Sci. Instrum.}\ }\textbf {\bibinfo {volume} {55}},\ \bibinfo
  {pages} {231--234} (\bibinfo {year} {1984})},\ \Eprint
  {http://arxiv.org/abs/https://doi.org/10.1063/1.1137730}
  {https://doi.org/10.1063/1.1137730} \BibitemShut {NoStop}%
\bibitem [{\citenamefont {Torikachvili}\ \emph {et~al.}(2015)\citenamefont
  {Torikachvili}, \citenamefont {Kim}, \citenamefont {Colombier}, \citenamefont
  {Bud’ko},\ and\ \citenamefont {Canfield}}]{Torikachvili2015}%
  \BibitemOpen
  \bibfield  {author} {\bibinfo {author} {\bibfnamefont {M.~S.}\ \bibnamefont
  {Torikachvili}}, \bibinfo {author} {\bibfnamefont {S.~K.}\ \bibnamefont
  {Kim}}, \bibinfo {author} {\bibfnamefont {E.}~\bibnamefont {Colombier}},
  \bibinfo {author} {\bibfnamefont {S.~L.}\ \bibnamefont {Bud’ko}}, \ and\
  \bibinfo {author} {\bibfnamefont {P.~C.}\ \bibnamefont {Canfield}},\
  }\bibfield  {title} {\enquote {\bibinfo {title} {Solidification and loss of
  hydrostaticity in liquid media used for pressure measurements},}\ }\href
  {\doibase 10.1063/1.4937478} {\bibfield  {journal} {\bibinfo  {journal} {Rev.
  Sci. Instrum.}\ }\textbf {\bibinfo {volume} {86}},\ \bibinfo {pages} {123904}
  (\bibinfo {year} {2015})},\ \Eprint
  {http://arxiv.org/abs/http://dx.doi.org/10.1063/1.4937478}
  {http://dx.doi.org/10.1063/1.4937478} \BibitemShut {NoStop}%
\bibitem [{\citenamefont {Eiling}\ and\ \citenamefont
  {Schilling}(1981)}]{Eiling1981}%
  \BibitemOpen
  \bibfield  {author} {\bibinfo {author} {\bibfnamefont {A}~\bibnamefont
  {Eiling}}\ and\ \bibinfo {author} {\bibfnamefont {J~S}\ \bibnamefont
  {Schilling}},\ }\bibfield  {title} {\enquote {\bibinfo {title} {{Pressure and
  temperature dependence of electrical resistivity of Pb and Sn from 1-300K and
  0-10 GPa-use as continuous resistive pressure monitor accurate over wide
  temperature range; superconductivity under pressure in Pb, Sn and In}},}\
  }\href {http://stacks.iop.org/0305-4608/11/i=3/a=010} {\bibfield  {journal}
  {\bibinfo  {journal} {J. Phys. F: Met. Phys}\ }\textbf {\bibinfo {volume}
  {11}},\ \bibinfo {pages} {623} (\bibinfo {year} {1981})}\BibitemShut
  {NoStop}%
\bibitem [{\citenamefont {Bireckoven}\ and\ \citenamefont
  {Wittig}(1988)}]{Bireckoven1988}%
  \BibitemOpen
  \bibfield  {author} {\bibinfo {author} {\bibfnamefont {B}~\bibnamefont
  {Bireckoven}}\ and\ \bibinfo {author} {\bibfnamefont {J}~\bibnamefont
  {Wittig}},\ }\bibfield  {title} {\enquote {\bibinfo {title} {{A diamond anvil
  cell for the investigation of superconductivity under pressures of up to 50
  GPa: Pb as a low temperature manometer}},}\ }\href
  {http://stacks.iop.org/0022-3735/21/i=9/a=004} {\bibfield  {journal}
  {\bibinfo  {journal} {J Phys E}\ }\textbf {\bibinfo {volume} {21}},\ \bibinfo
  {pages} {841} (\bibinfo {year} {1988})}\BibitemShut {NoStop}%
\bibitem [{\citenamefont {Kim}\ \emph {et~al.}(2011)\citenamefont {Kim},
  \citenamefont {Torikachvili}, \citenamefont {Colombier}, \citenamefont
  {Thaler}, \citenamefont {Bud'ko},\ and\ \citenamefont {Canfield}}]{Kim2011}%
  \BibitemOpen
  \bibfield  {author} {\bibinfo {author} {\bibfnamefont {S.~K.}\ \bibnamefont
  {Kim}}, \bibinfo {author} {\bibfnamefont {M.~S.}\ \bibnamefont
  {Torikachvili}}, \bibinfo {author} {\bibfnamefont {E.}~\bibnamefont
  {Colombier}}, \bibinfo {author} {\bibfnamefont {A.}~\bibnamefont {Thaler}},
  \bibinfo {author} {\bibfnamefont {S.~L.}\ \bibnamefont {Bud'ko}}, \ and\
  \bibinfo {author} {\bibfnamefont {P.~C.}\ \bibnamefont {Canfield}},\
  }\bibfield  {title} {\enquote {\bibinfo {title} {{Combined effects of
  pressure and Ru substitution on BaFe${}_{2}$As${}_{2}$}},}\ }\href {\doibase
  10.1103/PhysRevB.84.134525} {\bibfield  {journal} {\bibinfo  {journal} {Phys.
  Rev. B}\ }\textbf {\bibinfo {volume} {84}},\ \bibinfo {pages} {134525}
  (\bibinfo {year} {2011})}\BibitemShut {NoStop}%
\bibitem [{SHE()}]{SHELXTL}%
  \BibitemOpen
  \href@noop {} {\emph {\bibinfo {title} {SHELXTL-v2008/4, Bruker AXS Inc.,
  Madison, Wisconsin, USA, 2013.}}}\BibitemShut {Stop}%
\bibitem [{\citenamefont {Toby}(2001)}]{Toby:hw0089}%
  \BibitemOpen
  \bibfield  {author} {\bibinfo {author} {\bibfnamefont {Brian~H.}\
  \bibnamefont {Toby}},\ }\bibfield  {title} {\enquote {\bibinfo {title} {{{\it
  EXPGUI}, a graphical user interface for {\it GSAS}}},}\ }\href {\doibase
  10.1107/S0021889801002242} {\bibfield  {journal} {\bibinfo  {journal} {J.
  Appl. Crystallogr.}\ }\textbf {\bibinfo {volume} {34}},\ \bibinfo {pages}
  {210--213} (\bibinfo {year} {2001})}\BibitemShut {NoStop}%
\bibitem [{\citenamefont {Sucksmith}\ and\ \citenamefont
  {Thompson}(1954)}]{Sucksmith362}%
  \BibitemOpen
  \bibfield  {author} {\bibinfo {author} {\bibfnamefont {W.}~\bibnamefont
  {Sucksmith}}\ and\ \bibinfo {author} {\bibfnamefont {J.~E.}\ \bibnamefont
  {Thompson}},\ }\bibfield  {title} {\enquote {\bibinfo {title} {{The Magnetic
  Anisotropy of Cobalt}},}\ }\href {\doibase 10.1098/rspa.1954.0209} {\bibfield
   {journal} {\bibinfo  {journal} {Proc. Royal Soc. A}\ }\textbf {\bibinfo
  {volume} {225}},\ \bibinfo {pages} {362--375} (\bibinfo {year}
  {1954})}\BibitemShut {NoStop}%
\bibitem [{\citenamefont {Lamichhane}\ \emph
  {et~al.}(2016{\natexlab{a}})\citenamefont {Lamichhane}, \citenamefont
  {Taufour}, \citenamefont {Masters}, \citenamefont {Parker}, \citenamefont
  {Kaluarachchi}, \citenamefont {Thimmaiah}, \citenamefont {Bud'ko},\ and\
  \citenamefont {Canfield}}]{HfZrMnPTej}%
  \BibitemOpen
  \bibfield  {author} {\bibinfo {author} {\bibfnamefont {Tej~N.}\ \bibnamefont
  {Lamichhane}}, \bibinfo {author} {\bibfnamefont {Valentin}\ \bibnamefont
  {Taufour}}, \bibinfo {author} {\bibfnamefont {Morgan~W.}\ \bibnamefont
  {Masters}}, \bibinfo {author} {\bibfnamefont {David~S.}\ \bibnamefont
  {Parker}}, \bibinfo {author} {\bibfnamefont {Udhara~S.}\ \bibnamefont
  {Kaluarachchi}}, \bibinfo {author} {\bibfnamefont {Srinivasa}\ \bibnamefont
  {Thimmaiah}}, \bibinfo {author} {\bibfnamefont {Sergey~L.}\ \bibnamefont
  {Bud'ko}}, \ and\ \bibinfo {author} {\bibfnamefont {Paul~C.}\ \bibnamefont
  {Canfield}},\ }\bibfield  {title} {\enquote {\bibinfo {title} {Discovery of
  ferromagnetism with large magnetic anisotropy in zrmnp and hfmnp},}\ }\href
  {\doibase 10.1063/1.4961933} {\bibfield  {journal} {\bibinfo  {journal}
  {Appl. Phys. Lett.}\ }\textbf {\bibinfo {volume} {109}},\ \bibinfo {pages}
  {092402} (\bibinfo {year} {2016}{\natexlab{a}})},\ \Eprint
  {http://arxiv.org/abs/http://dx.doi.org/10.1063/1.4961933}
  {http://dx.doi.org/10.1063/1.4961933} \BibitemShut {NoStop}%
\bibitem [{\citenamefont {Md~Din}\ \emph {et~al.}(2015)\citenamefont {Md~Din},
  \citenamefont {Wang}, \citenamefont {Cheng}, \citenamefont {Dou},
  \citenamefont {Kennedy}, \citenamefont {Avdeev},\ and\ \citenamefont
  {Campbell}}]{MdDin2015}%
  \BibitemOpen
  \bibfield  {author} {\bibinfo {author} {\bibfnamefont {M.~F.}\ \bibnamefont
  {Md~Din}}, \bibinfo {author} {\bibfnamefont {J.~L.}\ \bibnamefont {Wang}},
  \bibinfo {author} {\bibfnamefont {Z.~X.}\ \bibnamefont {Cheng}}, \bibinfo
  {author} {\bibfnamefont {S.~X.}\ \bibnamefont {Dou}}, \bibinfo {author}
  {\bibfnamefont {S.~J.}\ \bibnamefont {Kennedy}}, \bibinfo {author}
  {\bibfnamefont {M.}~\bibnamefont {Avdeev}}, \ and\ \bibinfo {author}
  {\bibfnamefont {S.~J.}\ \bibnamefont {Campbell}},\ }\bibfield  {title}
  {\enquote {\bibinfo {title} {{Tuneable Magnetic Phase Transitions in Layered
  CeMn$_2$Ge$_{2-x}$Si$_x$ Compounds}},}\ }\href {\doibase 10.1038/srep11288}
  {\bibfield  {journal} {\bibinfo  {journal} {‎Sci. Rep}\ }\textbf {\bibinfo
  {volume} {5}} (\bibinfo {year} {2015}),\ 10.1038/srep11288}\BibitemShut
  {NoStop}%
\bibitem [{\citenamefont {Wang}\ \emph {et~al.}(2009)\citenamefont {Wang},
  \citenamefont {Campbell}, \citenamefont {Zeng}, \citenamefont {Poh},
  \citenamefont {Dou},\ and\ \citenamefont {Kennedy}}]{JLWang2009}%
  \BibitemOpen
  \bibfield  {author} {\bibinfo {author} {\bibfnamefont {J.~L.}\ \bibnamefont
  {Wang}}, \bibinfo {author} {\bibfnamefont {S.~J.}\ \bibnamefont {Campbell}},
  \bibinfo {author} {\bibfnamefont {R.}~\bibnamefont {Zeng}}, \bibinfo {author}
  {\bibfnamefont {C.~K.}\ \bibnamefont {Poh}}, \bibinfo {author} {\bibfnamefont
  {S.~X.}\ \bibnamefont {Dou}}, \ and\ \bibinfo {author} {\bibfnamefont
  {S.~J.}\ \bibnamefont {Kennedy}},\ }\bibfield  {title} {\enquote {\bibinfo
  {title} {{Re-entrant ferromagnet PrMn$_2$Ge$_{0.8}$Si$_{1.2}$: Magnetocaloric
  effect}},}\ }\href {\doibase 10.1063/1.3059610} {\bibfield  {journal}
  {\bibinfo  {journal} {J. Appl. Phys.}\ }\textbf {\bibinfo {volume} {105}},\
  \bibinfo {pages} {07A909} (\bibinfo {year} {2009})},\ \Eprint
  {http://arxiv.org/abs/https://doi.org/10.1063/1.3059610}
  {https://doi.org/10.1063/1.3059610} \BibitemShut {NoStop}%
\bibitem [{\citenamefont {Chen}\ \emph {et~al.}(2010)\citenamefont {Chen},
  \citenamefont {Luo}, \citenamefont {Liang}, \citenamefont {Li},\ and\
  \citenamefont {Rao}}]{CHEN201013}%
  \BibitemOpen
  \bibfield  {author} {\bibinfo {author} {\bibfnamefont {Y.Q.}\ \bibnamefont
  {Chen}}, \bibinfo {author} {\bibfnamefont {J.}~\bibnamefont {Luo}}, \bibinfo
  {author} {\bibfnamefont {J.K.}\ \bibnamefont {Liang}}, \bibinfo {author}
  {\bibfnamefont {J.B.}\ \bibnamefont {Li}}, \ and\ \bibinfo {author}
  {\bibfnamefont {G.H.}\ \bibnamefont {Rao}},\ }\bibfield  {title} {\enquote
  {\bibinfo {title} {{Magnetic properties and magnetocaloric effect of
  Nd(Mn$_{1 - x}$Fe$_x$)$_2$Ge$_2$ compounds}},}\ }\href {\doibase
  https://doi.org/10.1016/j.jallcom.2009.09.078} {\bibfield  {journal}
  {\bibinfo  {journal} {J. Alloys Compd.}\ }\textbf {\bibinfo {volume} {489}},\
  \bibinfo {pages} {13 -- 19} (\bibinfo {year} {2010})}\BibitemShut {NoStop}%
\bibitem [{\citenamefont {Pecharsky}\ and\ \citenamefont
  {Gschneidner}(1997)}]{Pecharsky1997}%
  \BibitemOpen
  \bibfield  {author} {\bibinfo {author} {\bibfnamefont {V.~K.}\ \bibnamefont
  {Pecharsky}}\ and\ \bibinfo {author} {\bibfnamefont {K.~A.}\ \bibnamefont
  {Gschneidner}, \bibfnamefont {Jr.}},\ }\bibfield  {title} {\enquote {\bibinfo
  {title} {{Giant Magnetocaloric Effect in
  ${\mathrm{Gd}}_{5}({\mathrm{Si}}_{2}{\mathrm{Ge}}_{2})$}},}\ }\href {\doibase
  10.1103/PhysRevLett.78.4494} {\bibfield  {journal} {\bibinfo  {journal}
  {Phys. Rev. Lett.}\ }\textbf {\bibinfo {volume} {78}},\ \bibinfo {pages}
  {4494--4497} (\bibinfo {year} {1997})}\BibitemShut {NoStop}%
\bibitem [{\citenamefont {Lamichhane}\ \emph
  {et~al.}(2016{\natexlab{b}})\citenamefont {Lamichhane}, \citenamefont
  {Taufour}, \citenamefont {Thimmaiah}, \citenamefont {Parker}, \citenamefont
  {Bud'ko},\ and\ \citenamefont {Canfield}}]{Lamichhane2015}%
  \BibitemOpen
  \bibfield  {author} {\bibinfo {author} {\bibfnamefont {Tej~N.}\ \bibnamefont
  {Lamichhane}}, \bibinfo {author} {\bibfnamefont {Valentin}\ \bibnamefont
  {Taufour}}, \bibinfo {author} {\bibfnamefont {Srinivasa}\ \bibnamefont
  {Thimmaiah}}, \bibinfo {author} {\bibfnamefont {David~S.}\ \bibnamefont
  {Parker}}, \bibinfo {author} {\bibfnamefont {Sergey~L.}\ \bibnamefont
  {Bud'ko}}, \ and\ \bibinfo {author} {\bibfnamefont {Paul~C.}\ \bibnamefont
  {Canfield}},\ }\bibfield  {title} {\enquote {\bibinfo {title} {{A study of
  the physical properties of single crystalline Fe$_5$B$_2$P}},}\ }\href
  {\doibase 10.1016/j.jmmm.2015.10.088} {\bibfield  {journal} {\bibinfo
  {journal} {J. Magn. Magn. Mater.}\ }\textbf {\bibinfo {volume} {401}},\
  \bibinfo {pages} {525 -- 531} (\bibinfo {year}
  {2016}{\natexlab{b}})}\BibitemShut {NoStop}%
\bibitem [{\citenamefont {Banerjee}(1964)}]{BANERJEE1964}%
  \BibitemOpen
  \bibfield  {author} {\bibinfo {author} {\bibfnamefont {B.K.}\ \bibnamefont
  {Banerjee}},\ }\bibfield  {title} {\enquote {\bibinfo {title} {On a
  generalised approach to first and second order magnetic transitions},}\
  }\href {\doibase http://dx.doi.org/10.1016/0031-9163(64)91158-8} {\bibfield
  {journal} {\bibinfo  {journal} {Phys. Lett.}\ }\textbf {\bibinfo {volume}
  {12}},\ \bibinfo {pages} {16 -- 17} (\bibinfo {year} {1964})}\BibitemShut
  {NoStop}%
\bibitem [{\citenamefont {Arrott}\ and\ \citenamefont
  {Noakes}(1967)}]{Arrott1967}%
  \BibitemOpen
  \bibfield  {author} {\bibinfo {author} {\bibfnamefont {Anthony}\ \bibnamefont
  {Arrott}}\ and\ \bibinfo {author} {\bibfnamefont {John~E.}\ \bibnamefont
  {Noakes}},\ }\bibfield  {title} {\enquote {\bibinfo {title} {Approximate
  equation of state for nickel near its critical temperature},}\ }\href
  {\doibase 10.1103/PhysRevLett.19.786} {\bibfield  {journal} {\bibinfo
  {journal} {Phys. Rev. Lett.}\ }\textbf {\bibinfo {volume} {19}},\ \bibinfo
  {pages} {786--789} (\bibinfo {year} {1967})}\BibitemShut {NoStop}%
\bibitem [{\citenamefont {Kouvel}\ and\ \citenamefont {Fisher}(1964)}]{KF1964}%
  \BibitemOpen
  \bibfield  {author} {\bibinfo {author} {\bibfnamefont {James~S.}\
  \bibnamefont {Kouvel}}\ and\ \bibinfo {author} {\bibfnamefont {Michael~E.}\
  \bibnamefont {Fisher}},\ }\bibfield  {title} {\enquote {\bibinfo {title}
  {Detailed magnetic behavior of nickel near its curie point},}\ }\href
  {\doibase 10.1103/PhysRev.136.A1626} {\bibfield  {journal} {\bibinfo
  {journal} {Phys. Rev.}\ }\textbf {\bibinfo {volume} {136}},\ \bibinfo {pages}
  {A1626--A1632} (\bibinfo {year} {1964})}\BibitemShut {NoStop}%
\bibitem [{\citenamefont {Mohan}\ \emph {et~al.}(1998)\citenamefont {Mohan},
  \citenamefont {Seeger}, \citenamefont {Kronmüller}, \citenamefont
  {Murugaraj},\ and\ \citenamefont {Maier}}]{MohanKFLaSrMNO3}%
  \BibitemOpen
  \bibfield  {author} {\bibinfo {author} {\bibfnamefont {Ch.V}\ \bibnamefont
  {Mohan}}, \bibinfo {author} {\bibfnamefont {M}~\bibnamefont {Seeger}},
  \bibinfo {author} {\bibfnamefont {H}~\bibnamefont {Kronmüller}}, \bibinfo
  {author} {\bibfnamefont {P}~\bibnamefont {Murugaraj}}, \ and\ \bibinfo
  {author} {\bibfnamefont {J}~\bibnamefont {Maier}},\ }\bibfield  {title}
  {\enquote {\bibinfo {title} {{Critical behaviour near the
  ferromagnetic–paramagnetic phase transition in
  La$_{0.8}$Sr$_{0.2}$MnO$_{3}$ }},}\ }\href {\doibase
  https://doi.org/10.1016/S0304-8853(97)01095-0} {\bibfield  {journal}
  {\bibinfo  {journal} {J. Magn. Magn. Mater}\ }\textbf {\bibinfo {volume}
  {183}},\ \bibinfo {pages} {348 -- 355} (\bibinfo {year} {1998})}\BibitemShut
  {NoStop}%
\bibitem [{\citenamefont {Chen}\ \emph {et~al.}(2013)\citenamefont {Chen},
  \citenamefont {Yang}, \citenamefont {Wang}, \citenamefont {Imai},
  \citenamefont {Ohta}, \citenamefont {Michioka}, \citenamefont {Yoshimura},\
  and\ \citenamefont {Fang}}]{BinChenFe3GeTe2}%
  \BibitemOpen
  \bibfield  {author} {\bibinfo {author} {\bibfnamefont {Bin}\ \bibnamefont
  {Chen}}, \bibinfo {author} {\bibfnamefont {JinHu}\ \bibnamefont {Yang}},
  \bibinfo {author} {\bibfnamefont {HangDong}\ \bibnamefont {Wang}}, \bibinfo
  {author} {\bibfnamefont {Masaki}\ \bibnamefont {Imai}}, \bibinfo {author}
  {\bibfnamefont {Hiroto}\ \bibnamefont {Ohta}}, \bibinfo {author}
  {\bibfnamefont {Chishiro}\ \bibnamefont {Michioka}}, \bibinfo {author}
  {\bibfnamefont {Kazuyoshi}\ \bibnamefont {Yoshimura}}, \ and\ \bibinfo
  {author} {\bibfnamefont {MingHu}\ \bibnamefont {Fang}},\ }\bibfield  {title}
  {\enquote {\bibinfo {title} {{Magnetic Properties of Layered Itinerant
  Electron Ferromagnet Fe$_3$GeTe$_2$}},}\ }\href {\doibase
  10.7566/JPSJ.82.124711} {\bibfield  {journal} {\bibinfo  {journal} {J. Phys.
  Soc. Jpn.}\ }\textbf {\bibinfo {volume} {82}},\ \bibinfo {pages} {124711}
  (\bibinfo {year} {2013})},\ \Eprint
  {http://arxiv.org/abs/http://dx.doi.org/10.7566/JPSJ.82.124711}
  {http://dx.doi.org/10.7566/JPSJ.82.124711} \BibitemShut {NoStop}%
\bibitem [{\citenamefont {Pramanik}\ and\ \citenamefont
  {Banerjee}(2009)}]{ParamanikandBanarjee}%
  \BibitemOpen
  \bibfield  {author} {\bibinfo {author} {\bibfnamefont {A.~K.}\ \bibnamefont
  {Pramanik}}\ and\ \bibinfo {author} {\bibfnamefont {A.}~\bibnamefont
  {Banerjee}},\ }\bibfield  {title} {\enquote {\bibinfo {title} {{Critical
  behavior at paramagnetic to ferromagnetic phase transition in
  ${\text{Pr}}_{0.5}{\text{Sr}}_{0.5}{\text{MnO}}_{3}$: A bulk magnetization
  study}},}\ }\href {\doibase 10.1103/PhysRevB.79.214426} {\bibfield  {journal}
  {\bibinfo  {journal} {Phys. Rev. B}\ }\textbf {\bibinfo {volume} {79}},\
  \bibinfo {pages} {214426} (\bibinfo {year} {2009})}\BibitemShut {NoStop}%
\bibitem [{\citenamefont {Rhodes}\ and\ \citenamefont
  {Wohlfarth}(1963)}]{RWratio1963}%
  \BibitemOpen
  \bibfield  {author} {\bibinfo {author} {\bibfnamefont {P.}~\bibnamefont
  {Rhodes}}\ and\ \bibinfo {author} {\bibfnamefont {E.~P.}\ \bibnamefont
  {Wohlfarth}},\ }\bibfield  {title} {\enquote {\bibinfo {title} {The effective
  curie-weiss constant of ferromagnetic metals and alloys},}\ }\href {\doibase
  10.1098/rspa.1963.0086} {\bibfield  {journal} {\bibinfo  {journal} {Proc.
  Roy. Soc. A. Mathematical, Physical and Engineering Sciences}\ } (\bibinfo
  {year} {1963}),\ 10.1098/rspa.1963.0086}\BibitemShut {NoStop}%
\bibitem [{\citenamefont {Pfleiderer}\ \emph {et~al.}(2004)\citenamefont
  {Pfleiderer}, \citenamefont {Reznik}, \citenamefont {Pintschovius},
  \citenamefont {v.~L\"ohneysen}, \citenamefont {Garst},\ and\ \citenamefont
  {Rosch}}]{Pfleiderer2004}%
  \BibitemOpen
  \bibfield  {author} {\bibinfo {author} {\bibfnamefont {C.}~\bibnamefont
  {Pfleiderer}}, \bibinfo {author} {\bibfnamefont {D.}~\bibnamefont {Reznik}},
  \bibinfo {author} {\bibfnamefont {L.}~\bibnamefont {Pintschovius}}, \bibinfo
  {author} {\bibfnamefont {H.}~\bibnamefont {v.~L\"ohneysen}}, \bibinfo
  {author} {\bibfnamefont {M.}~\bibnamefont {Garst}}, \ and\ \bibinfo {author}
  {\bibfnamefont {A.}~\bibnamefont {Rosch}},\ }\bibfield  {title} {\enquote
  {\bibinfo {title} {{Partial order in the non-Fermi-liquid phase of MnSi}},}\
  }\href {\doibase 10.1038/nature02232} {\bibfield  {journal} {\bibinfo
  {journal} {Nature}\ }\textbf {\bibinfo {volume} {427}},\ \bibinfo {pages}
  {227–231} (\bibinfo {year} {2004})}\BibitemShut {NoStop}%
\bibitem [{\citenamefont {Uhlarz}\ \emph {et~al.}(2004)\citenamefont {Uhlarz},
  \citenamefont {Pfleiderer},\ and\ \citenamefont {Hayden}}]{Uhlarz2004}%
  \BibitemOpen
  \bibfield  {author} {\bibinfo {author} {\bibfnamefont {M.}~\bibnamefont
  {Uhlarz}}, \bibinfo {author} {\bibfnamefont {C.}~\bibnamefont {Pfleiderer}},
  \ and\ \bibinfo {author} {\bibfnamefont {S.~M.}\ \bibnamefont {Hayden}},\
  }\bibfield  {title} {\enquote {\bibinfo {title} {{Quantum Phase Transitions
  in the Itinerant Ferromagnet
  ${\mathrm{Z}\mathrm{r}\mathrm{Z}\mathrm{n}}_{2}$}},}\ }\href {\doibase
  10.1103/PhysRevLett.93.256404} {\bibfield  {journal} {\bibinfo  {journal}
  {Phys. Rev. Lett.}\ }\textbf {\bibinfo {volume} {93}},\ \bibinfo {pages}
  {256404} (\bibinfo {year} {2004})}\BibitemShut {NoStop}%
\bibitem [{\citenamefont {Niklowitz}\ \emph {et~al.}(2005)\citenamefont
  {Niklowitz}, \citenamefont {Beckers}, \citenamefont {Lonzarich},
  \citenamefont {Knebel}, \citenamefont {Salce}, \citenamefont {Thomasson},
  \citenamefont {Bernhoeft}, \citenamefont {Braithwaite},\ and\ \citenamefont
  {Flouquet}}]{Niklowitz2005}%
  \BibitemOpen
  \bibfield  {author} {\bibinfo {author} {\bibfnamefont {P.~G.}\ \bibnamefont
  {Niklowitz}}, \bibinfo {author} {\bibfnamefont {F.}~\bibnamefont {Beckers}},
  \bibinfo {author} {\bibfnamefont {G.~G.}\ \bibnamefont {Lonzarich}}, \bibinfo
  {author} {\bibfnamefont {G.}~\bibnamefont {Knebel}}, \bibinfo {author}
  {\bibfnamefont {B.}~\bibnamefont {Salce}}, \bibinfo {author} {\bibfnamefont
  {J.}~\bibnamefont {Thomasson}}, \bibinfo {author} {\bibfnamefont
  {N.}~\bibnamefont {Bernhoeft}}, \bibinfo {author} {\bibfnamefont
  {D.}~\bibnamefont {Braithwaite}}, \ and\ \bibinfo {author} {\bibfnamefont
  {J.}~\bibnamefont {Flouquet}},\ }\bibfield  {title} {\enquote {\bibinfo
  {title} {{Spin-fluctuation-dominated electrical transport of
  ${\mathrm{Ni}}_{3}\mathrm{Al}$ at high pressure}},}\ }\href {\doibase
  10.1103/PhysRevB.72.024424} {\bibfield  {journal} {\bibinfo  {journal} {Phys.
  Rev. B}\ }\textbf {\bibinfo {volume} {72}},\ \bibinfo {pages} {024424}
  (\bibinfo {year} {2005})}\BibitemShut {NoStop}%
\bibitem [{\citenamefont {Blaha}\ \emph {et~al.}(2001)\citenamefont {Blaha},
  \citenamefont {Schwarz}, \citenamefont {Madsen}, \citenamefont {Kvasnicka},\
  and\ \citenamefont {Luitz}}]{wien2k}%
  \BibitemOpen
  \bibfield  {author} {\bibinfo {author} {\bibfnamefont {P.}~\bibnamefont
  {Blaha}}, \bibinfo {author} {\bibfnamefont {K.}~\bibnamefont {Schwarz}},
  \bibinfo {author} {\bibfnamefont {G.~K.~H.}\ \bibnamefont {Madsen}}, \bibinfo
  {author} {\bibfnamefont {D.}~\bibnamefont {Kvasnicka}}, \ and\ \bibinfo
  {author} {\bibfnamefont {J.}~\bibnamefont {Luitz}},\ }\href@noop {} {\emph
  {\bibinfo {title} {{WIEN2K}, {A}n {A}ugmented {P}lane {W}ave + {L}ocal
  {O}rbitals {P}rogram for {C}alculating {C}rystal {P}roperties}}}\ (\bibinfo
  {publisher} {{K}arlheinz Schwarz, Techn. Universit\"{a}t Wien, Austria},\
  \bibinfo {year} {2001})\BibitemShut {NoStop}%
\bibitem [{\citenamefont {Sjostedt}\ \emph {et~al.}(2000)\citenamefont
  {Sjostedt}, \citenamefont {Nordstrom},\ and\ \citenamefont
  {Singh}}]{singh_lo}%
  \BibitemOpen
  \bibfield  {author} {\bibinfo {author} {\bibfnamefont {E}~\bibnamefont
  {Sjostedt}}, \bibinfo {author} {\bibfnamefont {L}~\bibnamefont {Nordstrom}},
  \ and\ \bibinfo {author} {\bibfnamefont {David.~J}\ \bibnamefont {Singh}},\
  }\bibfield  {title} {\enquote {\bibinfo {title} {An alternative way of
  linearizing the augmented plane-wave method},}\ }\href {\doibase
  10.1016/S0038-1098(99)00577-3} {\bibfield  {journal} {\bibinfo  {journal}
  {Solid State Commun.}\ }\textbf {\bibinfo {volume} {114}},\ \bibinfo {pages}
  {15 -- 20} (\bibinfo {year} {2000})}\BibitemShut {NoStop}%
\bibitem [{\citenamefont {Singh}\ and\ \citenamefont
  {Nordstrom}(2006)}]{singh_book}%
  \BibitemOpen
  \bibfield  {author} {\bibinfo {author} {\bibfnamefont {David~J.}\
  \bibnamefont {Singh}}\ and\ \bibinfo {author} {\bibfnamefont
  {L.}~\bibnamefont {Nordstrom}},\ }\href {\doibase 10.1007/978-0-387-29684-5}
  {\emph {\bibinfo {title} {Planewaves Pseudopotentials and the LAPW Method,
  2nd ed.}}}\ (\bibinfo  {publisher} {Springer},\ \bibinfo {address} {Berlin},\
  \bibinfo {year} {2006})\BibitemShut {NoStop}%
\bibitem [{\citenamefont {Perdew}\ \emph {et~al.}(1996)\citenamefont {Perdew},
  \citenamefont {Burke},\ and\ \citenamefont
  {Ernzerhof}}]{perdew1996generalized}%
  \BibitemOpen
  \bibfield  {author} {\bibinfo {author} {\bibfnamefont {John~P}\ \bibnamefont
  {Perdew}}, \bibinfo {author} {\bibfnamefont {Kieron}\ \bibnamefont {Burke}},
  \ and\ \bibinfo {author} {\bibfnamefont {Matthias}\ \bibnamefont
  {Ernzerhof}},\ }\bibfield  {title} {\enquote {\bibinfo {title} {Generalized
  gradient approximation made simple},}\ }\href {\doibase
  10.1103/PhysRevLett.77.3865} {\bibfield  {journal} {\bibinfo  {journal}
  {Phys. Rev. Lett.}\ }\textbf {\bibinfo {volume} {77}},\ \bibinfo {pages}
  {3865} (\bibinfo {year} {1996})}\BibitemShut {NoStop}%
\bibitem [{\citenamefont {Cedervall}\ \emph
  {et~al.}(2016{\natexlab{b}})\citenamefont {Cedervall}, \citenamefont
  {Andersson}, \citenamefont {Sarkar}, \citenamefont {Delczeg-Czirjak},
  \citenamefont {Bergqvist}, \citenamefont {Hansen}, \citenamefont {Beran},
  \citenamefont {Nordblad},\ and\ \citenamefont
  {Sahlberg}}]{cedervall2016magnetic}%
  \BibitemOpen
  \bibfield  {author} {\bibinfo {author} {\bibfnamefont {Johan}\ \bibnamefont
  {Cedervall}}, \bibinfo {author} {\bibfnamefont {Mikael~Svante}\ \bibnamefont
  {Andersson}}, \bibinfo {author} {\bibfnamefont {Tapati}\ \bibnamefont
  {Sarkar}}, \bibinfo {author} {\bibfnamefont {Erna~K}\ \bibnamefont
  {Delczeg-Czirjak}}, \bibinfo {author} {\bibfnamefont {Lars}\ \bibnamefont
  {Bergqvist}}, \bibinfo {author} {\bibfnamefont {Thomas~C}\ \bibnamefont
  {Hansen}}, \bibinfo {author} {\bibfnamefont {Premysl}\ \bibnamefont {Beran}},
  \bibinfo {author} {\bibfnamefont {Per}\ \bibnamefont {Nordblad}}, \ and\
  \bibinfo {author} {\bibfnamefont {Martin}\ \bibnamefont {Sahlberg}},\
  }\bibfield  {title} {\enquote {\bibinfo {title} {{Magnetic structure of the
  magnetocaloric compound AlFe$_2$B$_2$}},}\ }\href {\doibase
  10.1016/j.jallcom.2015.12.111} {\bibfield  {journal} {\bibinfo  {journal} {J.
  Alloys Compd.}\ }\textbf {\bibinfo {volume} {664}},\ \bibinfo {pages}
  {784--791} (\bibinfo {year} {2016}{\natexlab{b}})}\BibitemShut {NoStop}%
\bibitem [{\citenamefont {Palmer}(1963)}]{palmer1963magnetocrystalline}%
  \BibitemOpen
  \bibfield  {author} {\bibinfo {author} {\bibfnamefont {Wilfred}\ \bibnamefont
  {Palmer}},\ }\bibfield  {title} {\enquote {\bibinfo {title}
  {Magnetocrystalline anisotropy of magnetite at low temperature},}\ }\href
  {\doibase 10.1103/PhysRev.131.1057} {\bibfield  {journal} {\bibinfo
  {journal} {Phys. Rev.}\ }\textbf {\bibinfo {volume} {131}},\ \bibinfo {pages}
  {1057} (\bibinfo {year} {1963})}\BibitemShut {NoStop}%
\bibitem [{\citenamefont {Coey}(2011)}]{coey2011hard}%
  \BibitemOpen
  \bibfield  {author} {\bibinfo {author} {\bibfnamefont {J.~M.~D.}\
  \bibnamefont {Coey}},\ }\bibfield  {title} {\enquote {\bibinfo {title} {Hard
  magnetic materials: A perspective},}\ }\href {\doibase
  10.1109/TMAG.2011.2166975} {\bibfield  {journal} {\bibinfo  {journal} {IEEE
  Trans. Magn.}\ }\textbf {\bibinfo {volume} {47}},\ \bibinfo {pages}
  {4671--4681} (\bibinfo {year} {2011})}\BibitemShut {NoStop}%
\end{thebibliography}%
\end{document}